\documentclass[acmtog]{acmart}
\copyrightyear{2026}
\acmYear{2026}
\setcopyright{cc}
\setcctype{by}
\acmConference[SA Conference Papers '26]{SIGGRAPH Asia 2026 Conference Papers}{December 01--04, 2026}{Kuala Lumpur, Malaysia}
\acmBooktitle{SIGGRAPH Asia 2026 Conference Papers (SA Conference Papers '26), December 01--04, 2026, Kuala Lumpur, Malaysia}
\acmDOI{10.1145/3829340.3842344}
\acmISBN{979-8-4007-2842-6/2026/12}

\usepackage[ruled]{algorithm2e} 

\SetAlFnt{\small}
\SetAlCapFnt{\small}
\SetAlCapNameFnt{\small}
\SetAlCapHSkip{0pt}

\makeatletter
\renewcommand*{\@fnsymbol}[1]{\ifcase#1\or \TextOrMath\textdagger\dagger
  \or \TextOrMath\textdaggerdbl\ddagger \or \TextOrMath\textsection\mathsection
  \or \TextOrMath\textparagraph\mathparagraph \else\@ctrerr\fi}
\makeatother
\begin{document}
\title{SCRIPT: Scalable Diffusion Policy with Multi-stage Training for Language-driven Physics-Based Humanoid Control}
\author{Jingyan Zhang}
\orcid{0009-0007-1970-2693}
\affiliation{%
 \institution{ShanghaiTech University}
 \city{Shanghai}
 \country{China}}
\email{zhangjy7@shanghaitech.edu.cn}

\author{Han Liang}
\authornote{Project leader.} 
\orcid{0009-0008-2444-9685}
\affiliation{%
 \institution{ByteDance Seed}
  \city{Shanghai}
 \country{China}}
\email{lianghan@shanghaitech.edu.cn}

\author{Ruichi Zhang}
\orcid{0009-0009-2503-4999}
\affiliation{%
 \institution{University of Pennsylvania}
 \city{Philadelphia}
 \country{USA}}
\email{rczhang@seas.upenn.edu}

\author{Bin Li}
\orcid{0009-0005-5072-1708}
\affiliation{%
 \institution{ShanghaiTech University}
  \city{Shanghai}
 \country{China}}
\email{libin3@shanghaitech.edu.cn}

\author{Juze Zhang}
\orcid{0000-0003-0583-2722}
\affiliation{%
 \institution{Stanford University}
 \city{Stanford}
 \country{USA}}
\email{juze@stanford.edu}

\author{Xin Chen}
\orcid{0000-0002-9347-1367}
\affiliation{%
 \institution{ByteDance Seed}
 \city{San Jose}
 \country{USA}}
\email{chenxin2@shanghaitech.edu.cn}

\author{Jingya Wang}
\orcid{0000-0003-1122-0634}
\affiliation{%
 \institution{ShanghaiTech University}
 \city{Shanghai}
 \country{China}
}
\email{wangjingya@shanghaitech.edu.cn}

\author{Lan Xu}
\orcid{0000-0002-8807-7787}
\affiliation{%
 \institution{ShanghaiTech University}
 \city{Shanghai}
 \country{China}
}
\email{xulan1@shanghaitech.edu.cn}

\author{Jingyi Yu}
\authornote{Corresponding author.}
\orcid{0000-0002-8580-0036}
\affiliation{%
 \institution{ShanghaiTech University}
 \city{Shanghai}
 \country{China}
}
\email{yujingyi@shanghaitech.edu.cn}
\renewcommand{\shortauthors}{Zhang et al.}
\begin{abstract}
Controlling physics-based humanoids from natural-language instructions is a critical step toward general-purpose embodied agents.
However, existing methods remain constrained by a tension between semantic expressiveness and physical feasibility, often failing to jointly achieve faithful instruction following, high-quality motion, and stable long-horizon control.
We propose SCRIPT, a scalable diffusion policy with a multi-stage training framework for language-driven physics-based humanoid control.
The core of SCRIPT is a Joint Action-State-Text Diffusion Transformer (JAST-DiT), which represents actions, physical states, and text as dedicated token streams and couples them through joint attention, enabling direct interaction between language semantics and control dynamics.
To stabilize autoregressive control, we introduce a nonlinear history conditioning mechanism, which preserves the dense recent context and samples increasingly sparse cues from long-term history.
Beyond supervised imitation pre-training, we propose a post-training stage, further improving the performance using Reinforcement Learning with Hybrid Rewards (RLHR). 
By injecting learnable noise into the flow-sampling process, 
RLHR effectively improves motion quality and instruction following within closed-loop simulations using hybrid physical feedback and text rewards.
Quantitative evaluations demonstrate that SCRIPT outperforms prior state-of-the-art methods, with gains across text alignment, motion quality, and physical realism metrics. 
Furthermore, scaling studies on the 1200-hour MotionMillion dataset demonstrate consistent performance gains with model scaling, highlighting SCRIPT's robust scalability for large-scale pre-training.
Code and data for this paper are available at \url{https://zhanglele12138.github.io/SCRIPT/}.
\end{abstract}

%
%
%
%
\begin{teaserfigure}
  \centering
  \includegraphics[width=\textwidth]{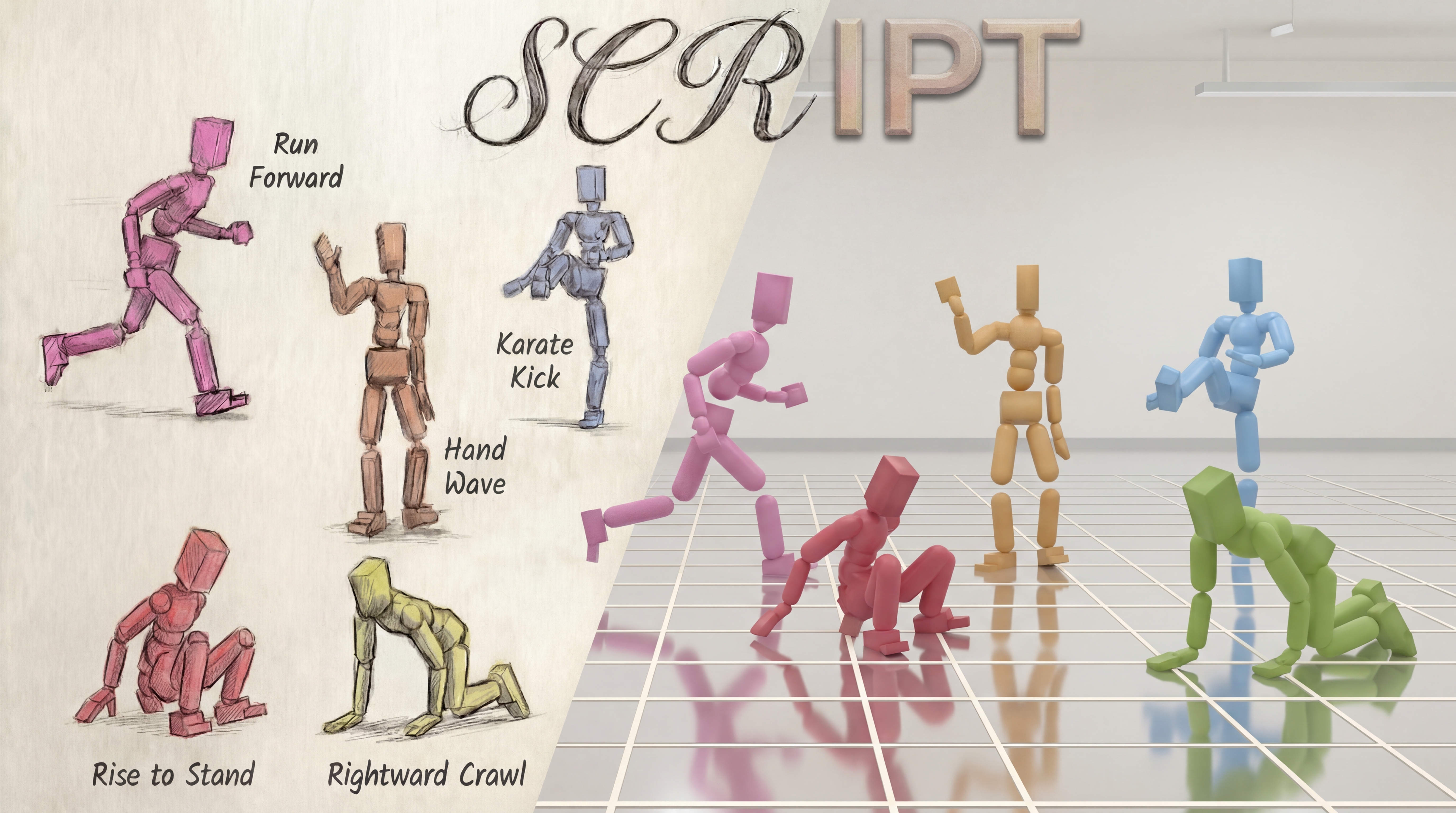}
  \caption{
    SCRIPT translates natural-language motion descriptions (left) into physically simulated humanoid behavior (right) under closed-loop dynamics.
  }
    \Description{Text prompts on the left paired with physically simulated humanoid motions on the right, each showing a character performing the described action in the simulator.}
  \label{fig:teaser}
\end{teaserfigure}

\maketitle
\section{Introduction}
Endowing humanoid agents with the ability to comprehend human intent and interact naturally with their environments is pivotal to bridging the digital and physical worlds. 
Toward this vision, enabling physics-based humanoids to synthesize expressive and physically plausible motions from language instructions has emerged as a key frontier in computer graphics and embodied artificial intelligence~\cite{fung2025embodied}.
Achieving such capability at scale fundamentally hinges on massive paired text-motion data~\cite{Plappert2016,mahmood2019amass,guo2022generating}.
Recent advances in large language models (LLMs)~\cite{brown2020language,touvron2023llama,guo2025deepseek} and vision-language models (VLMs)~\cite{radford2021learning,liu2023visual,bai2025qwen3} have made it feasible to construct richly annotated, large-scale motion datasets from diverse web sources~\cite{lin2023motion,zhang2025motion,fan2025go,mclean2025embody}.
The expansion of data scale and semantic coverage has fueled rapid progress in text-conditioned kinematic motion generation, enabling models to synthesize diverse human motion sequences from open-vocabulary instructions~\cite{liang2024omg,jiang2023motiongpt,tevet2022human,fan2025go,lu2025scamo,wen2025hy}. 
In parallel, deep reinforcement learning (RL) has become a powerful paradigm for physics-based character control, where policies are optimized through interaction with simulated environments. By training policies directly under rigid-body dynamics in contact-rich simulations, these methods can produce controllers for highly dynamic skills such as running, agile traversal, and acrobatic maneuvers~\cite{peng2021amp,peng2022ase,rempe2023trace,zhang2025physics,xu2025parc}.
Despite these advances, combining the scalability of data-driven kinematic motion generation with the physical feasibility and closed-loop robustness of humanoid control remains challenging.

Recent studies~\cite{ren2023insactor,serifi2024robot,tevet2024closd,wu2025human} adopt a hierarchical framework, where diffusion-based planners generate kinematic references and a tracking policy executes them under physical constraints. However, since the references are not guaranteed to be dynamically feasible, the tracker must trade off reference fidelity against actuation limits, and subtle reference artifacts can be amplified into visible failures during closed-loop execution.
Other approaches build unified controllers by distilling specialized expert policies~\cite{juravsky2024superpadl} or leveraging pretrained motion priors~\cite{mu2025smp}. 
Despite advancing policy coverage and instruction following, compressing diverse motions into a single policy incurs information loss and mode averaging, compromising stylistic diversity~\cite{peng2022ase,zhu2023neural}.
Recent large-scale tracking policies show that RL-based humanoid control can benefit from scaling, but they remain focused on reference-motion tracking rather than language-conditioned control~\cite{luo2025sonic}.
Diffusion policies have also been explored for physics-based humanoid control, offering expressive action distributions for closed-loop execution~\cite{ren2024diffusion,truong2024pdp,huang2025diffuse,wu2025uniphys}. However, by treating text merely as a coarse condition, these methods weakly couple language with states and actions, limiting joint improvements in text-motion alignment and physical plausibility.

Building on these insights, we propose SCRIPT, a scalable end-to-end diffusion policy with a multi-stage training paradigm for language-driven physics-based humanoid control. 
Conditioned on language instructions and state histories, SCRIPT generates executable actions to drive humanoid motion within a physics simulator. 
SCRIPT is built upon the Joint Action-State-Text Diffusion Transformer (JAST-DiT), which maintains separate token streams for actions, physical states, and text while enabling dense cross-modal interaction through joint attention.
To support stable autoregressive execution, we further introduce a nonlinear long-term history conditioning mechanism that preserves recent control dynamics and sparsely samples distant context during rollouts.
Beyond imitation pre-training with supervised flow matching loss, SCRIPT incorporates an RL post-training stage that directly optimizes task-level rewards to further improve semantic alignment and physical stability. Experiments show that SCRIPT achieves state-of-the-art performance on HumanML3D and exhibits consistent scaling gains on MotionMillion as the model size increases from 0.2B to 1.2B parameters.
Our contributions are summarized as follows:
\begin{itemize}
\item {We introduce SCRIPT, a scalable diffusion policy framework for language-driven physics-based humanoid control, achieving state-of-the-art performance.}
\item {We propose JAST-DiT and a tailored nonlinear long-term history conditioning mechanism for stable autoregressive modeling, demonstrating favorable scaling performance on the large-scale 1200-hour MotionMillion dataset.}
\item {We propose a hybrid-reward RL post-training stage that optimizes physics-based and text-alignment rewards, further improving physical stability and instruction following.}
\end{itemize}

\begin{figure*}[t]
  \centering
  \includegraphics[width=0.88\textwidth]{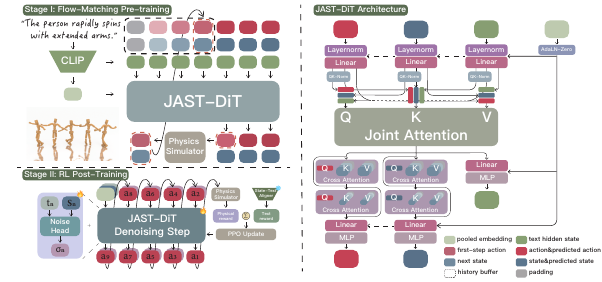}
  \caption{Overview of the SCRIPT framework. Left: Stage I pre-trains a flow matching diffusion policy via behavior cloning, and Stage II applies RL post-training via PPO with hybrid rewards. JAST-DiT jointly models action, state, and text tokens through separate streams with joint attention.}
  \label{fig:method_overview}
    \Description{Two-panel diagram: the left panel shows Stage I behavior-cloning pre-training followed by Stage II PPO post-training with hybrid rewards; the right panel shows the JAST-DiT block, where action, state, and text token streams interact through joint attention.}
\end{figure*}
\section{Related Work}
\paragraph{Large-Scale Kinematic Motion Generation.}
Human motion datasets have rapidly expanded in both scale and modality coverage, spanning text~\cite{guo2022generating,zhang2025motion,fan2025go}, audio~\cite{li2021learn, liu2022beat, liu2024emage}, inertial signals~\cite{DIP:SIGGRAPHAsia:2018,Trumble:BMVC:2017}, and objects~\cite{bhatnagar2022behave,li2023object,zhang2024hoi}. 
Advances in LLM- and VLM-assisted annotation have made it increasingly feasible to construct large-scale paired text-motion datasets.
Driven by the steady growth of text-motion data, recent models have increasingly adopted scalable generative architectures~\cite{peebles2023scalable,jiang2023motiongpt,wang2024motiongpt,zhu2025motiongpt3,he2025molingo}.
By scaling up model capacity, recent works have advanced motion generation from closed-set to open-vocabulary generation, achieving substantially improved text-motion alignment for complex textual prompts~\cite{liang2024omg,lu2025scamo, fan2025go,rempe2026kimodo,li2026llamo,wen2025hy}.
However, these methods generate offline kinematic trajectories rather than closed-loop control policies executable in physics simulation. Without modeling of contact dynamics, balance control, actuation limits, and environmental feedback, the resulting motions often exhibit physical artifacts such as foot skating, floating, and ground penetration, limiting their direct applicability to physics-based control.

\paragraph{Physics-Based Humanoid Control.}
Early methods in physics-based humanoid control largely followed a tracking paradigm~\cite{peng2018deepmimic}. Subsequent work introduced motion priors~\cite{peng2021amp,luo2024universal} and latent skill spaces~\cite{peng2022ase,tessler2023calm,yao2022controlvae,yao2024moconvq} to support reusable motor skills, while more recent language-conditioned controllers~\cite{juravsky2022padl,juravsky2024superpadl,tessler2024maskedmimic,mu2025smp} further enable text-driven control. 
Although effective in scaling, distilling diverse motions into compact policies or latent skill spaces often introduces mode averaging.
Another line of work follows a cascaded planner-tracker paradigm, where kinematic references are tracked by a physics-based controller~\cite{ren2023insactor,lim2026bric,tevet2024closd}. 
However, artifacts and infeasible contacts in the generated kinematic references can be amplified during tracking, degrading the physical stability and instruction fidelity.
Recent works apply diffusion policies to humanoid control~\cite{truong2024pdp,huang2025diffuse,wu2025uniphys}, locomotion for real robots~\cite{huang2024diffuseloco}, and motion tracking~\cite{liao2025beyondmimic}, training end-to-end controllers via behavior cloning on expert demonstrations. 
These methods typically rely on basic commands or reference trajectories, and even approaches incorporating language suffer from limited data scale and coarse conditioning that hinder precise alignment with physical control.
Building scalable language-driven diffusion policies from large-scale motion data while preserving physical plausibility and instruction fidelity remains an open challenge.

\paragraph{Reinforcement Learning for Diffusion Policies.}
Supervised diffusion training learns policies through denoising or flow matching regression, whereas RL fine-tunes policies through reward-driven environmental interaction.
To bridge this gap, recent methods formulate the denoising process as a Markov decision process (MDP) to optimize specific rewards~\cite{black2023training,fan2023dpok,yang2024using,hiranaka2024hero,hu2025towards,xue2025dancegrpo}.
This paradigm naturally extends to continuous control, where online RL is used to improve diffusion models initialized through behavioral cloning~\cite{ren2024diffusion,zhang2025reinflow}.
Policy gradients tailored to flow-based policies have also been applied to robot control~\cite{yi2026flow}.
Parallel efforts in human motion generation apply RL or preference optimization to align diffusion models with physical constraints, text prompts, and human aesthetics~\cite{han2024reindiffuse,liu2024motionrl,pappa2024modipo,girolamo2026no}.
However, these methods primarily focus on offline kinematic generation, where rewards are evaluated on completed trajectories rather than through closed-loop physical interaction. 
By contrast, physical humanoid control requires continuous execution, where each action shapes future states, causing minor deviations to compound into physical instability.
Therefore, RL post-training of diffusion-based policies that achieves both high-level text-motion alignment and stable closed-loop physical execution remains largely underexplored. 
Our RLHR complements these efforts by coupling a hybrid semantic-physical reward with closed-loop humanoid control.

\section{Preliminaries}
\label{sec:preliminaries}
\subsection{Problem Formulation}
\label{sec:problem_formulation}
Given a text instruction $c$, we train a language-driven diffusion policy $\pi_\theta$ to control a physics-based humanoid in a simulated environment.
Let $\mathbf{s}_t \in \mathbb{R}^{d_s}$ and $\mathbf{a}_t \in \mathbb{R}^{d_a}$ denote the proprioceptive state and the control action at time step $t$.
Let $L$ and $H$ denote the sampled state history length and the prediction horizon, respectively.
We define the sampled state history at time $t$ as 
$\mathcal{H}_t = \{\mathbf{s}_{t-\ell_i}\}_{i=1}^{L}$, 
where $\ell_i$ is the temporal offset selected from a raw history window by our history sampler.
The state $\mathbf{s}_t$ includes the root height, along with the local position, 6D rotation~\cite{zhou2019continuity}, linear velocity, and angular velocity of each joint. The action $\mathbf{a}_t$ specifies target joint angles for a proportional-derivative (PD) controller.
We encode the text instruction $c$ using the CLIP~\cite{radford2021learning} text encoder, yielding a pooled embedding $\mathbf{c}_{\text{pool}}$ and penultimate-layer token features $\mathbf{c}_{\text{txt}}$. 
Inspired by~\cite{esser2024scaling}, we utilize penultimate-layer features to prevent the loss of token-level semantics caused by final projections.
At each rollout step, $\pi_\theta$ jointly generates the next $H$ state-action pairs conditioned on the sampled state history $\mathcal{H}_t$ and the text embeddings:
\begin{equation}
  \pi_\theta\!\left(
  \mathbf{a}_{t+1:t+H},\,
  \mathbf{s}_{t+1:t+H}
  \mid
  \mathcal{H}_t,\,
  \mathbf{c}_{\text{pool}},\,
  \mathbf{c}_{\text{txt}}
  \right).
  \label{eq:policy}
\end{equation}
During inference, the policy operates in a receding-horizon manner, predicting an $H$-step trajectory, executing only the first action, and updating the history for the next prediction.

\subsection{Flow Matching}
\label{sec:flow_matching}
Flow Matching (FM) is a generalized formulation of diffusion models that replaces complex stochastic differential equations with simpler, deterministic probability paths. Accordingly, we formulate $\pi_\theta$ as a conditional generative policy based on FM~\cite{lipman2022flow,liu2022flow,albergo2022building}.
Let $\mathbf{x}_0 \sim \mathcal{N}(\mathbf{0}, \mathbf{I})$ denote a Gaussian noise sample and $\mathbf{x}_1 \sim p_{\mathrm{data}}$ denote a clean data sample. 
FM defines a linear probability path
\begin{equation}
  \mathbf{x}_\tau = (1-\tau) \mathbf{x}_0 + \tau \mathbf{x}_1,
  \qquad \tau \in [0,1],
  \label{eq:fm_interp}
\end{equation}
where $\tau$ is the flow time. 
The target velocity along this path is obtained by differentiating $\mathbf{x}_\tau$ with respect to $\tau$:
\begin{equation}
  \mathbf{u}_\tau =
  \frac{\mathrm{d}\mathbf{x}_\tau}{\mathrm{d}\tau}
  =
  \mathbf{x}_1 - \mathbf{x}_0.
  \label{eq:fm_velocity}
\end{equation}
FM trains a vector field 
$v_\theta(\mathbf{x}_\tau,\tau,\mathbf{y})$ to match this target velocity, where 
$\mathbf{y}=\{\mathcal{H}_t,\mathbf{c}_{\text{txt}},\mathbf{c}_{\text{pool}}\}$ denotes the conditioning information. 
The training objective is to minimize the mean squared error between the predicted and target velocities:
\begin{equation}
  \mathcal{L}_{\mathrm{FM}}(\theta)
  =
  \mathbb{E}_{\tau,\mathbf{x}_0,\mathbf{x}_1,\mathbf{y}}
  \left[
  \left\|
  v_\theta(\mathbf{x}_\tau,\tau,\mathbf{y})
  -
  \mathbf{u}_\tau
  \right\|_2^2
  \right].
  \label{eq:fm_loss}
\end{equation}
In our setting, the clean data sample $\mathbf{x}_1$ corresponds to the future state-action trajectory chunk
$[\mathbf{a}_{t+1:t+H}; \mathbf{s}_{t+1:t+H}]$.
\section{Methodology}
\label{sec:method}
Fig.~\ref{fig:method_overview} illustrates SCRIPT, our multi-stage framework for language-driven physics-based humanoid control. SCRIPT consists of three key components: 
(1) processing large-scale text-motion datasets to construct state-action demonstrations (Sec.~\ref{sec:data_curation}); (2) 
JAST-DiT, a diffusion transformer that jointly models actions, states, and text to predict future chunks, with nonlinear history-conditioned cross-attention (Sec.~\ref{sec:jast_dit}); 
and (3) an RLHR post-training stage that optimizes semantic alignment and physical stability via PPO~\cite{schulman2017proximal} on stochastic flow-sampling trajectories (Sec.~\ref{sec:rl_post_training}).
\subsection{Large-Scale Data Curation}
\label{sec:data_curation}
Large-scale physically executable data are essential for language-driven humanoid control. Existing text-motion datasets, such as HumanML3D~\cite{guo2022generating} and MotionMillion~\cite{fan2025go}, provide rich semantic coverage but consist of kinematic SMPL trajectories that often contain artifacts such as ground penetration and floating. Directly tracking these unconstrained motions in simulation can conflict with rigid-body dynamics, resulting in invalid state-action demonstrations.
To mitigate this issue, we curate the raw motions through pre-tracking artifact filtering and post-tracking rollout filtering with PHC~\cite{luo2023perpetual}. 
We filter out motion clips with severe artifacts, such as ground penetration or body floating.
Subsequently, we train multiple expert policies to track the remaining motions in parallel, collecting per-frame proprioceptive state $\mathbf{s}_t \in \mathbb{R}^{358}$ and target PD control action $\mathbf{a}_t \in \mathbb{R}^{69}$. 
Sequences with excessive tracking error or humanoid falls are discarded to avoid physically invalid demonstrations.
After curation, we retain 550K physical trajectories covering nearly 1200 hours of motion data, which are partitioned into training, validation, and test sets without motion overlap.
We use sliding windows to extract fixed-length segments from variable-length trajectories, with details provided in Sec.~A of the supplementary material.

\subsection{JAST-DiT Architecture}
\label{sec:jast_dit}
\paragraph{Joint Action-State-Text Self-Attention.}
To model the coupled dependencies among low-level actions, physical states, and language semantics, we introduce JAST-DiT, a Joint Action-State-Text Diffusion Transformer. 
Rather than using text solely as a global condition, our architecture maintains independent token streams for each modality, enabling modality-specific representation and token-level cross-modal interaction within each Transformer block.
As illustrated in the right panel of Fig.~\ref{fig:method_overview}, the inputs to JAST-DiT consist of the action and state tokens from the noisy trajectory chunk, alongside the penultimate-layer CLIP text tokens $\mathbf{c}_{\mathrm{txt}}$.
These tokens are mapped to a shared hidden dimension via stream-specific projections and subsequently augmented with positional embeddings.
We denote the resulting streams at the $\ell$-th Transformer block as $\mathbf{A}^{(\ell)}$, $\mathbf{S}^{(\ell)}$, and $\mathbf{C}^{(\ell)}$.
In parallel, the flow time $\tau$ and the pooled text embedding $\mathbf{c}_{\mathrm{pool}}$ are encoded into a global condition vector $\mathbf{e}=\mathrm{Cond}(\tau,\mathbf{c}_{\mathrm{pool}})$.
Within each block, this vector is injected via AdaLN-Zero modulation~\cite{peebles2023scalable} to independently modulate each stream before projecting them into stream-specific queries, keys, and values (e.g., $\mathbf{Q}_a, \mathbf{K}_a, \mathbf{V}_a$ for the action stream).
To enhance numerical stability, we apply QK-Norm~\cite{henry2020query,dehghani2023scaling} to the query and key features.
These stream-specific projections are subsequently concatenated along the token dimension to form $\mathbf{Q}=[\mathbf{Q}_a;\mathbf{Q}_s;\mathbf{Q}_c]$, with $\mathbf{K}$ and $\mathbf{V}$ constructed analogously. 
Since the diffusion policy denoises the entire trajectory chunk jointly, the action and state tokens attend bidirectionally without causal masking, while the text padding mask is retained. 
Finally, we compute joint attention over the concatenated sequence and partition the output back into the three distinct feature streams:
\begin{equation}
    \left[ \bar{\mathbf{A}}^{(\ell)}; \bar{\mathbf{S}}^{(\ell)}; \bar{\mathbf{C}}^{(\ell)} \right] = \mathrm{Attention} \left( \mathbf{Q}, \mathbf{K}, \mathbf{V} \right).
    \label{eq:joint_attention}
\end{equation}
\begin{figure}[t]
    \centering
    \includegraphics[width=\linewidth]{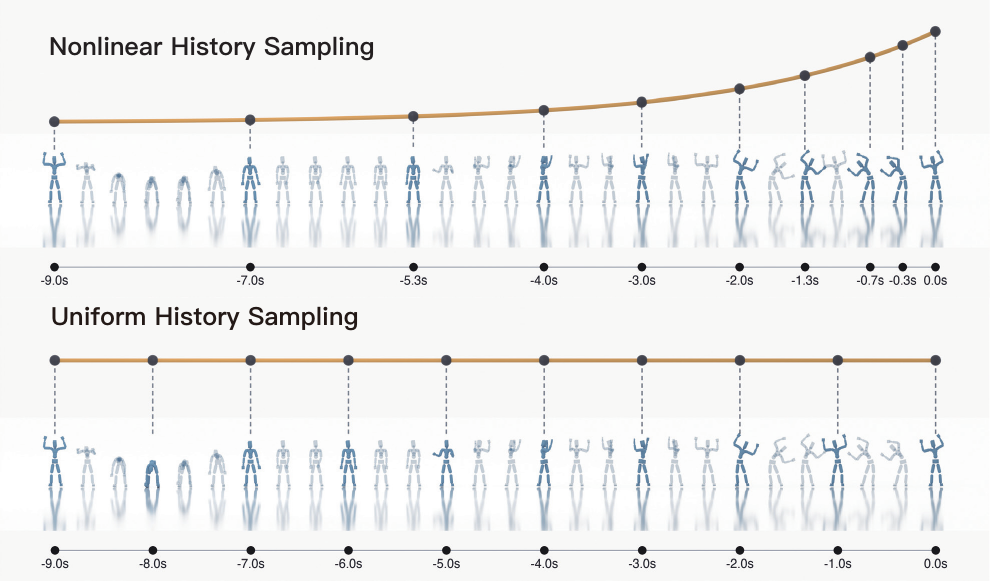}
    \caption{
    Nonlinear history sampling. Our strategy keeps recent states densely sampled and distant states sparsely sampled, balancing short-term control dynamics with long-range temporal context.
    }
    \Description{Timeline of past states in which sample points lie densely near the current frame and become progressively sparser further back in time.}
    \label{fig:sampling}
\end{figure}
\paragraph{Nonlinear History-Conditioned Cross-Attention.}
In the physical control of embodied agents, explicit memory mechanisms are crucial for maintaining long-term action coherence~\cite{fung2025embodied,radosavovic2024real,guhur2023instruction}. 
Conditioning solely on the current state provides limited temporal context, making it difficult to infer evolving dynamics such as momentum trends and contact transitions, which may lead to action drift or physical instability during long-horizon autoregressive rollouts.
Inspired by working memory~\cite{baddeley2020working} and forgetting curves~\cite{murre2015replication} in cognitive science, we incorporate short- and long-term state history into the action and state streams through cross-attention.
During rollouts, we maintain a dense history buffer $\mathcal{B}_t = \{(\mathbf{s}_{t-i})\}_{i=1}^{L_{\mathrm{max}}}$.
While the updated action and state streams from the joint attention layer could intuitively query this entire buffer, attending to all $L_{\mathrm{max}}$ historical states is computationally prohibitive and introduces redundant information.
To address this, we construct the sampled state history $\mathcal{H}_t$ using a nonlinear downsampling strategy, as illustrated in Fig.~\ref{fig:sampling}.
Specifically, we keep the latest $N_s$ states as dense recent history and sparsely sample $N_l$ frames from the remaining distant history of length $L_{\mathrm{distant}} = L_{\max} - N_s$.
To sample the distant history, we draw uniformly distributed variables $u_i \in [0,1]$ for $i=1,\ldots,N_l$ and compute the frame indices as:
\begin{equation}
    I_i = \left\lfloor L_{\mathrm{distant}}
    \left(
    1 + \frac{\ln\left(1 - u_i(1 - e^{-\alpha})\right)}{\alpha}
    \right)
    \right\rfloor .
    \label{eq:history_sampling}
\end{equation}
Here, $\alpha$ controls the decay rate, with larger values biasing the sampling probability toward more recent frames.
Rather than concatenating the distant and recent frames into a single sequence, we inject them hierarchically via sequential cross-attention.
The action and state streams attend to the sparse distant history $\mathbf{H}_{\mathrm{dist}}^{(\ell)}$ to retrieve long-range context, before attending to the dense recent history $\mathbf{H}_{\mathrm{rec}}^{(\ell)}$ to capture fine-grained local dynamics:
\begin{equation}
\mathbf{Q}_{\mathrm{inter}}
=
\mathrm{CrossAttention}
\left(
\left[\bar{\mathbf{A}}^{(\ell)};\bar{\mathbf{S}}^{(\ell)}\right],
\mathbf{H}_{\mathrm{dist}}^{(\ell)},
\mathbf{H}_{\mathrm{dist}}^{(\ell)}
\right).
\label{eq:dist_history_cross_attention}
\end{equation}
\begin{equation}
\left[
\hat{\mathbf{A}}^{(\ell)};
\hat{\mathbf{S}}^{(\ell)}
\right]
=
\mathrm{CrossAttention}
\left(
\mathbf{Q}_{\mathrm{inter}},
\mathbf{H}_{\mathrm{rec}}^{(\ell)},
\mathbf{H}_{\mathrm{rec}}^{(\ell)}
\right).
\label{eq:rec_history_cross_attention}
\end{equation}
After all JAST-DiT blocks, the refined token streams are projected to predict the flow velocity field $v_\theta$ for the future state-action trajectory chunk.
The entire model is trained with the conditional FM objective in Eq.~\ref{eq:fm_loss}, where the clean sample corresponds to the future state-action chunk $[\mathbf{a}_{t+1:t+H};\mathbf{s}_{t+1:t+H}]$.

\paragraph{Scalable Model Variants.}
To study the impact of model scaling on policy performance in physics-based humanoid control, we follow the scaling principles of Diffusion Transformers (DiT)~\cite{peebles2023scalable} and instantiate the SCRIPT model across three distinct capacities: Base, Large, and Huge.
As summarized in Table~\ref{tab:model_scaling}, we systematically scale the number of network layers $n_{\mathrm{layers}}$, the hidden dimension $d_{\mathrm{model}}$, and the number of attention heads $n_{\mathrm{heads}}$, while keeping the per-head dimension $d_{\mathrm{head}}$ fixed at 128. 
Consequently, the total parameter count ranges from 206.31M to 1231.39M. 
By instantiating these variants, we provide an empirical exploration of scaling behaviors for physics-based humanoid control.
\begin{table}[t]
  \caption{SCRIPT model variants.}
  \Description{Table listing the Base, Large, and Huge variants with their numbers of layers, hidden dimensions, attention heads, and parameter counts from 206M to 1231M.}
  \label{tab:model_scaling}
  \centering
  \small
  \renewcommand{\arraystretch}{1.08}
  \begin{tabular*}{0.9\columnwidth}{@{\extracolsep{\fill}}lccccc@{}}
    \toprule
    Model
    & $n_{\mathrm{layers}}$
    & $d_{\mathrm{model}}$
    & $n_{\mathrm{heads}}$
    & $d_{\mathrm{head}}$
    & Params \\
    \midrule
    SCRIPT-Base & $8$  & $512$  & $4$ & $128$ & $206.31$M \\
    SCRIPT-Large & $10$ & $768$  & $6$ & $128$ & $577.97$M \\
    SCRIPT-Huge & $12$ & $1024$ & $8$ & $128$ & $1231.39$M \\
    \bottomrule
  \end{tabular*}
\end{table}

\begin{table*}[t]
    \caption{Comparison with physics-based humanoid control baselines on HumanML3D. We comprehensively evaluate the methods across text-motion alignment, motion quality, and physical stability. The best results are in \textbf{bold}.}
    \Description{Table comparing SCRIPT with physics-based baselines on HumanML3D across text-motion alignment, motion quality, and physical stability metrics, with the best value of each column in bold.}
  \label{tab:comparison}
  \centering
  \small
  \setlength{\tabcolsep}{4.0pt}
  \renewcommand{\arraystretch}{1.08}
  \begin{tabular}{lccc@{\hspace{8pt}}ccc@{\hspace{8pt}}ccc}
    \toprule
    Method
    & \multicolumn{3}{c}{R-precision}
    & \multicolumn{3}{c}{Motion Quality}
    & \multicolumn{3}{c}{Physics-based Metrics} \\
    \cmidrule(lr){2-4}
    \cmidrule(lr){5-7}
    \cmidrule(lr){8-10}
    & Top-1 $\uparrow$
    & Top-2 $\uparrow$
    & Top-3 $\uparrow$
    & FID $\downarrow$
    & MM Dist. $\downarrow$
    & Diversity $\rightarrow$
    & Floating $\downarrow$
    & Jerk $\downarrow$
    & Duration time $\uparrow$ \\
    \midrule
    Phys-GT
    & $0.651{\pm}.002$
    & $0.815{\pm}.001$
    & $0.882{\pm}.001$
    & $0.000{\pm}.000$
    & $1.700{\pm}.001$
    & $1.494{\pm}.008$
    & $17.49$
    & $2.941$ 
    & $100.00\%$ \\
    \midrule
    PDP~\cite{truong2024pdp}
    & $0.206{\pm}.002$
    & $0.324{\pm}.001$
    & $0.416{\pm}.002$
    & $1.536{\pm}.004$
    & $2.666{\pm}.003$
    & $1.335{\pm}.011$
    & $27.19$
    & $3.047$
    & $89.54\%$ \\
    UniPhys~\cite{wu2025uniphys}
    & $0.143{\pm}.004$
    & $0.242{\pm}.003$
    & $0.326{\pm}.003$
    & $0.487{\pm}.001$
    & $2.750{\pm}.001$
    & $1.447{\pm}.008$
    & $19.67$
    & $2.036$ 
    & $92.55\%$ \\
    CLoSD~\cite{tevet2024closd}
    & $0.370{\pm}.002$
    & $0.537{\pm}.002$
    & $0.641{\pm}.001$
    & $0.728{\pm}.002$
    & $2.291{\pm}.001$
    & $1.444{\pm}.008$
    & $20.71$
    & $2.767$ 
    & $94.81\%$ \\
    \midrule
    SCRIPT Stage I
    & $0.429{\pm}.003$
    & $0.595{\pm}.003$
    & $0.689{\pm}.001$
    & $0.203{\pm}.001$
    & $\mathbf{2.112}{\pm}.001$
    & $1.462{\pm}.007$
    & $17.85$
    & $1.723$ 
    & $97.67\%$\\
    SCRIPT Stage II
    & $\mathbf{0.435}{\pm}.002$
    & $\mathbf{0.599}{\pm}.002$
    & $\mathbf{0.693}{\pm}.001$
    & $\mathbf{0.164}{\pm}.001$
    & $2.123{\pm}.001$
    & $\mathbf{1.486}{\pm}.009$
    & $\mathbf{17.61}$
    & $\mathbf{1.706}$
    & $\mathbf{98.08\%}$ \\
    \bottomrule
  \end{tabular}
\end{table*}
\subsection{Online Reinforcement Learning Post-Training}
\label{sec:rl_post_training}
\paragraph{Stochastic Exploration via Noise Injection.}
Although the pretrained diffusion policy can successfully follow language instructions, supervised flow matching primarily fits the training distribution rather than directly optimizing task rewards under closed-loop physics simulation.
This limitation motivates RLHR, our online reinforcement learning post-training stage, which directly optimizes task rewards in physics simulation to improve semantic alignment and physical stability.
Following ReinFlow~\cite{zhang2025reinflow}, we transform the deterministic flow sampling process into a stochastic one by injecting Gaussian noise exclusively into the action subspace at each Euler integration step: 
\begin{equation}
\mathbf{x}_{k+1}
=
\mathbf{x}_k
+
v_\theta(\mathbf{x}_k,\tau_k,\mathbf{y})\Delta\tau
+
\mathbf{M}_a \odot \sigma_\phi(\tau_k,\mathcal{H}_t) \odot \boldsymbol{\epsilon}_k ,
\label{eq:stochastic_flow}
\end{equation}
where $\mathbf{x}_k$ denotes the intermediate noisy state-action trajectory chunk at the $k$-th Euler step.
Here, $\boldsymbol{\epsilon}_k \sim \mathcal{N}(0,\mathbf{I})$ represents the injected noise, $\sigma_\phi(\tau_k,\mathcal{H}_t)$ denotes a dynamic exploration scale predicted by a lightweight network conditioned on the flow time and state history, and $\mathbf{M}_a$ is a binary mask that restricts the perturbation to the action dimensions. 
This stochastic update turns the deterministic flow sampler into a Markov chain over Euler steps, thereby enabling tractable action-space transition likelihoods for policy-gradient optimization.
\paragraph{Reward Design.}  
Our reward design combines dense physical regularization with trajectory-level semantic alignment.
The physical term is a positive tracking bonus that stabilizes closed-loop execution by rewarding agreement with the reference motion distribution, while the semantic term encourages consistency with the language instruction.
We optimize the composite reward $r_t = w_{\mathrm{phys}} r_{\mathrm{phys}}(t) + w_{\mathrm{text}} r_{\mathrm{text}}(t)$, where $w_{\mathrm{phys}}$ and $w_{\mathrm{text}}$ are weighting coefficients for the physical and semantic components, respectively. 
The physical reward $r_{\mathrm{phys}}(t)$ rewards agreement between the simulated humanoid state and the reference motion:
\begin{equation}
r_{\mathrm{phys}}(t)
=
\sum_{m \in \{p,q,v,\omega\}}
w_m \exp\!\left(-k_m \lVert d_m(t) \rVert^2\right),
\label{eq:phys_reward}
\end{equation}
where $d_m(t)$ denotes the discrepancy between the simulated rollout and the reference kinematic data in terms of keypoint positions $p$, orientations $q$, linear velocities $v$, and angular velocities $\omega$. 
To encourage semantic alignment, we introduce a sparse trajectory-level state-text contrastive reward $r_{\mathrm{text}}$, which is evaluated exclusively at episode termination. 
To achieve this, we leverage a frozen state-text contrastive model consisting of a state trajectory encoder $f_m$ and a text encoder $f_t$, which was pre-trained to align physical state sequences with language descriptions.
For each environment $i$ within a training batch $\mathcal{B}$, we encode its full episodic state sequence $\mathbf{S}_i^{\mathrm{full}}$ and compute the cosine similarity to the text prompt $c_j$ as
$S_{ij}=\langle f_m(\mathbf{S}_i^{\mathrm{full}}), f_t(c_j)\rangle$.
The semantic reward is then defined as the log-normalized probability of the corresponding text prompt:
\begin{equation}
r_{\mathrm{text}}^{(i)}(t)
=
\log
\frac{\exp(S_{ii})}
{\sum_{j \in \mathcal{B}} \exp(S_{ij})}
\cdot
\mathbb{I}\!\left[t = T_{\mathrm{end}}^{(i)}\right].
\label{eq:text_reward}
\end{equation}

\begin{figure*}
  \centering
  \includegraphics[width=0.9\linewidth]{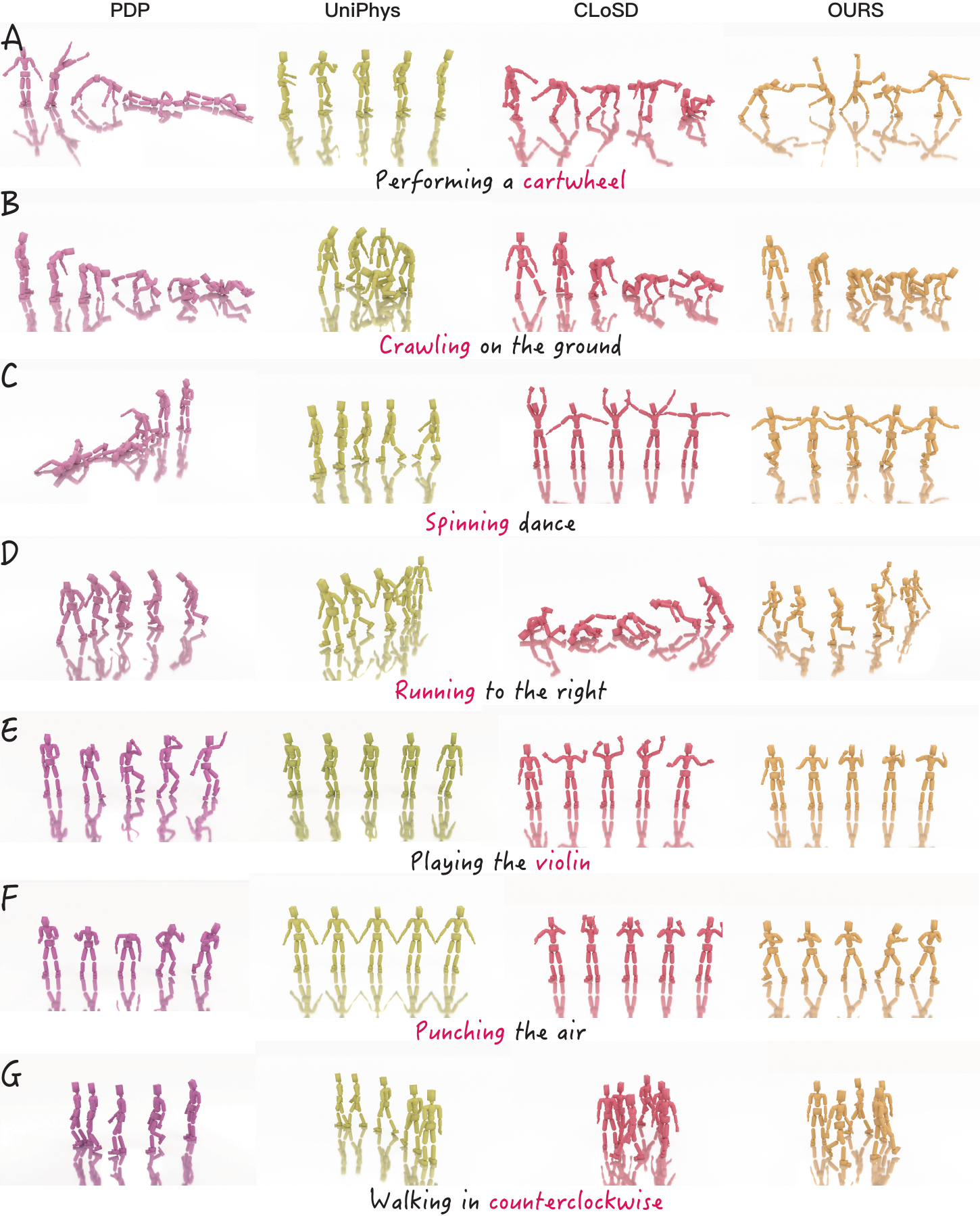}
  \caption{Qualitative comparison on HumanML3D. We compare SCRIPT against
PDP~\cite{truong2024pdp}, UniPhys~\cite{wu2025uniphys}, and CLoSD~\cite{tevet2024closd}. Each row shows a generated motion as a temporally stacked pose sequence, with the key prompt phrase highlighted. SCRIPT follows the prompt more faithfully while maintaining physical plausibility.}
  \Description{Grid of temporally stacked pose sequences comparing SCRIPT with PDP, UniPhys, and CLoSD on the same prompts; the highlighted phrase of each prompt is matched most closely by the SCRIPT row.}
  \label{fig:humanml3d-compare}
\end{figure*}
\paragraph{Optimization Objective.}
We optimize the stochastic flow sampler with PPO using the rewards defined above from physics rollouts.
To mitigate catastrophic forgetting and prevent excessive deviation from the pretrained flow matching policy, we augment the standard RL objective with a behavior cloning anchor loss:
\begin{equation}
\mathcal{L}_{\mathrm{RL}}
=
\mathcal{L}_{\mathrm{PPO}}
+
\alpha_v \mathcal{L}_{\mathrm{value}}
-
\alpha_e \mathcal{H}_{\mathrm{entropy}}
+
\alpha_{\mathrm{BC}} \mathcal{L}_{\mathrm{BC}} .
\label{eq:rl_objective}
\end{equation}
where $\mathcal{L}_{\mathrm{value}}$ represents the MSE between the critic predictions and the estimated returns, $\mathcal{H}_{\mathrm{entropy}}$ is the analytical action noise entropy.
To remain consistent with the first-stage flow matching objective, the BC anchor is evaluated on resampled flow matching pairs:
\begin{equation}
\mathcal{L}_{\mathrm{BC}}
=
\mathbb{E}_{\tau,\,\boldsymbol{\epsilon},\,\mathbf{x}}
\left[
\left\|
v_\theta(\mathbf{x}_\tau,\tau,\mathbf{y})
-
v_{\theta_0}(\mathbf{x}_\tau,\tau,\mathbf{y})
\right\|_2^2
\right],
\quad
\mathbf{x}_\tau=(1-\tau)\boldsymbol{\epsilon}+\tau\mathbf{x},
\label{eq:bc_loss}
\end{equation}
where $\theta_0$ denotes the frozen pretrained policy, and $\mathbf{x}$ is a clean state-action trajectory chunk sampled from the rollout buffer.
This term anchors the fine-tuned velocity field to the pretrained one while allowing PPO to improve task rewards in simulation. Further details are provided in Sec.~B of the supplementary material.
\section{Experiments}

\subsection{Experiment Setup}
\paragraph{Physics simulation setup.}
Following~\cite{luo2023perpetual}, we control a SMPL-like~\cite{loper2023smpl} physics humanoid in Isaac Gym simulator~\cite{makoviychuk2021isaac}. The humanoid consists of 24 body parts, with 23 actuated rotational joints and an unactuated pelvis. Given the current state $\mathbf{s}_t$ and action $\mathbf{a}_t$, the simulator advances the system as
$\mathbf{s}_{t+1} = \mathrm{SIM}(\mathbf{s}_t, \mathbf{a}_t)$.
\begin{figure*}
  \centering
  \includegraphics[width=0.9\linewidth]{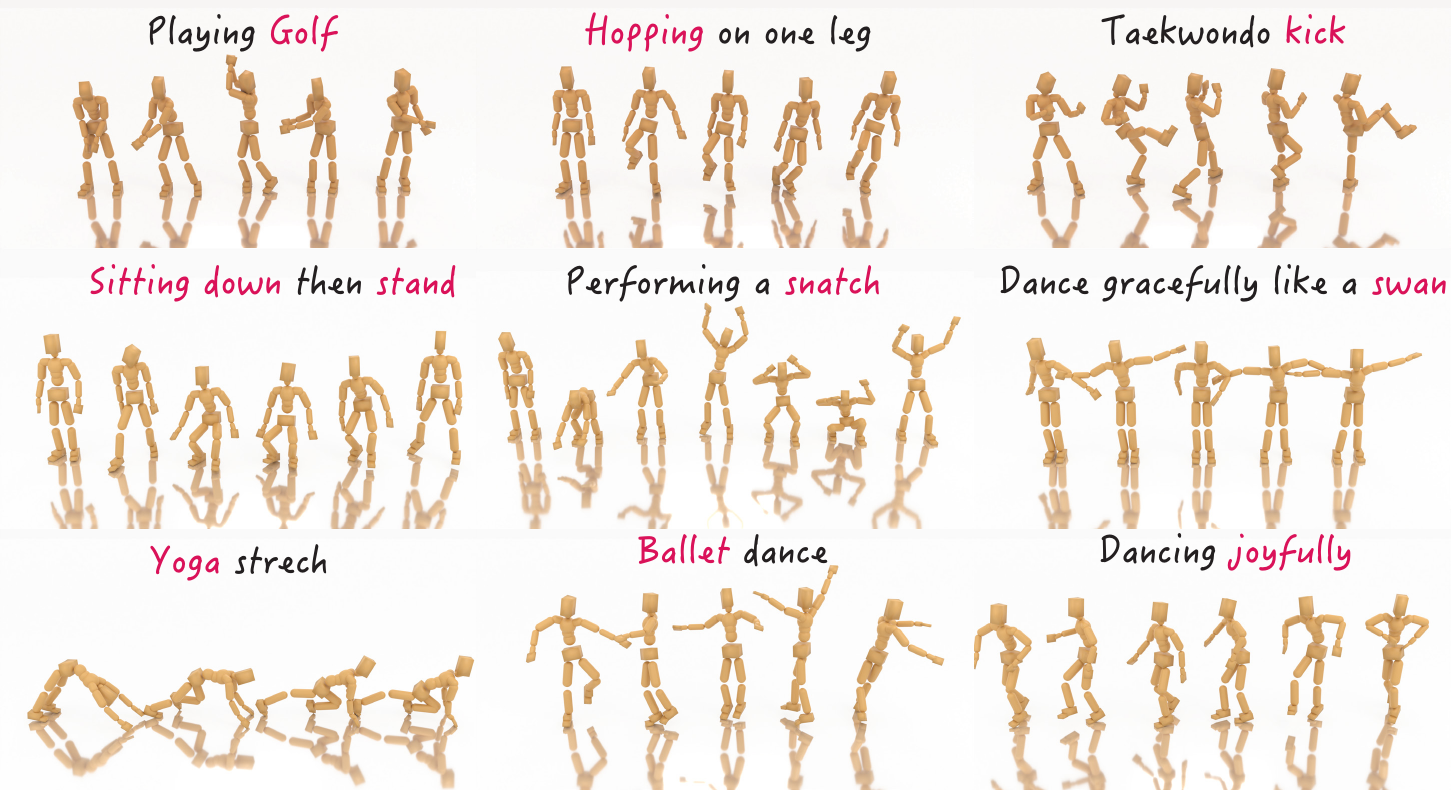}
  \caption{Qualitative results of SCRIPT-Huge trained on MotionMillion. Large-scale training enables diverse language-conditioned humanoid motions in physics simulation, covering locomotion, sports, dance, and daily actions.}
    \Description{Collage of humanoid motion sequences generated by SCRIPT-Huge in the physics simulator, spanning locomotion, sports, dance, and daily activities.}
  \label{fig:scaling}
\end{figure*}
\paragraph{Evaluation Metrics and Training Details.}
 We evaluate SCRIPT along two axes: text-motion alignment and physical plausibility. Following the text-to-motion evaluation protocol~\cite{guo2022generating,liang2024omg}, we report R-Precision (Top-1/2/3), FID, and MM-Dist using the text-motion evaluator. For physical plausibility, following physics-based motion benchmarks~\cite{yuan2023physdiff}, we report Floating, Jerk, and Duration time, where Floating and Jerk are computed from global joint positions. 
 SCRIPT is trained in two stages. 
 In Stage I, we train SCRIPT-Large on HumanML3D for benchmark comparisons and scaled model variants on MotionMillion for scaling studies. 
 The HumanML3D model uses $H=4$ and $L_{\max}=154$, while the MotionMillion scaling models use a longer history of $L_{\max}=604$. All models use 10\% text-condition dropout for classifier-free guidance~\cite{ho2022classifier}. 
 In Stage II, we initialize the actor from the pretrained model and perform RL post-training in 128 parallel Isaac Gym environments. 
 We use a BC anchor weight of 1.0 and constrain the learnable exploration noise to $\sigma\in[0.03,0.08]$. 
 Additional details are provided in the supplemental material.
\begin{table}[t]
  \caption{Model scaling on MotionMillion.}
  \Description{Table reporting evaluation metrics on MotionMillion for the Base, Large, and Huge models, showing steady improvement with model size.}
  \label{tab:scaling}
  \centering
  \small
  \renewcommand{\arraystretch}{1.08}
  \resizebox{\columnwidth}{!}{
  \begin{tabular}{lccc@{\hspace{8pt}}ccc}
    \toprule
    Method
    & \multicolumn{3}{c}{R-precision}
    & \multicolumn{3}{c}{Motion Quality} \\
    \cmidrule(lr){2-4}
    \cmidrule(lr){5-7}
    & Top-1 $\uparrow$
    & Top-2 $\uparrow$
    & Top-3 $\uparrow$
    & FID $\downarrow$
    & MM Dist. $\downarrow$
    & Diversity $\rightarrow$ \\
    \midrule
    GT
    & $0.707{\pm}.001$
    & $0.834{\pm}.001$
    & $0.886{\pm}.001$
    & $0.000{\pm}.000$
    & $2.864{\pm}.000$
    & $2.335{\pm}.018$ \\
    \midrule
    Base
    & $0.396{\pm}.001$
    & $0.544{\pm}.001$
    & $0.633{\pm}.001$
    & $1.057{\pm}.000$
    & $3.738{\pm}.000$
    & $2.251{\pm}.016$ \\
    Large
    & $0.437{\pm}.001$
    & $0.591{\pm}.001$
    & $0.680{\pm}.001$
    & $0.776{\pm}.000$
    & $3.625{\pm}.000$
    & $2.262{\pm}.012$ \\
    Huge
    & $\mathbf{0.464}{\pm}.001$
    & $\mathbf{0.616}{\pm}.001$
    & $\mathbf{0.701}{\pm}.001$
    & $\mathbf{0.645}{\pm}.000$
    & $\mathbf{3.554}{\pm}.000$
    & $\mathbf{2.287}{\pm}.020$ \\
    \bottomrule
  \end{tabular}
  }
\end{table}
\begin{figure*}
  \centering
  \includegraphics[width=0.9\linewidth]{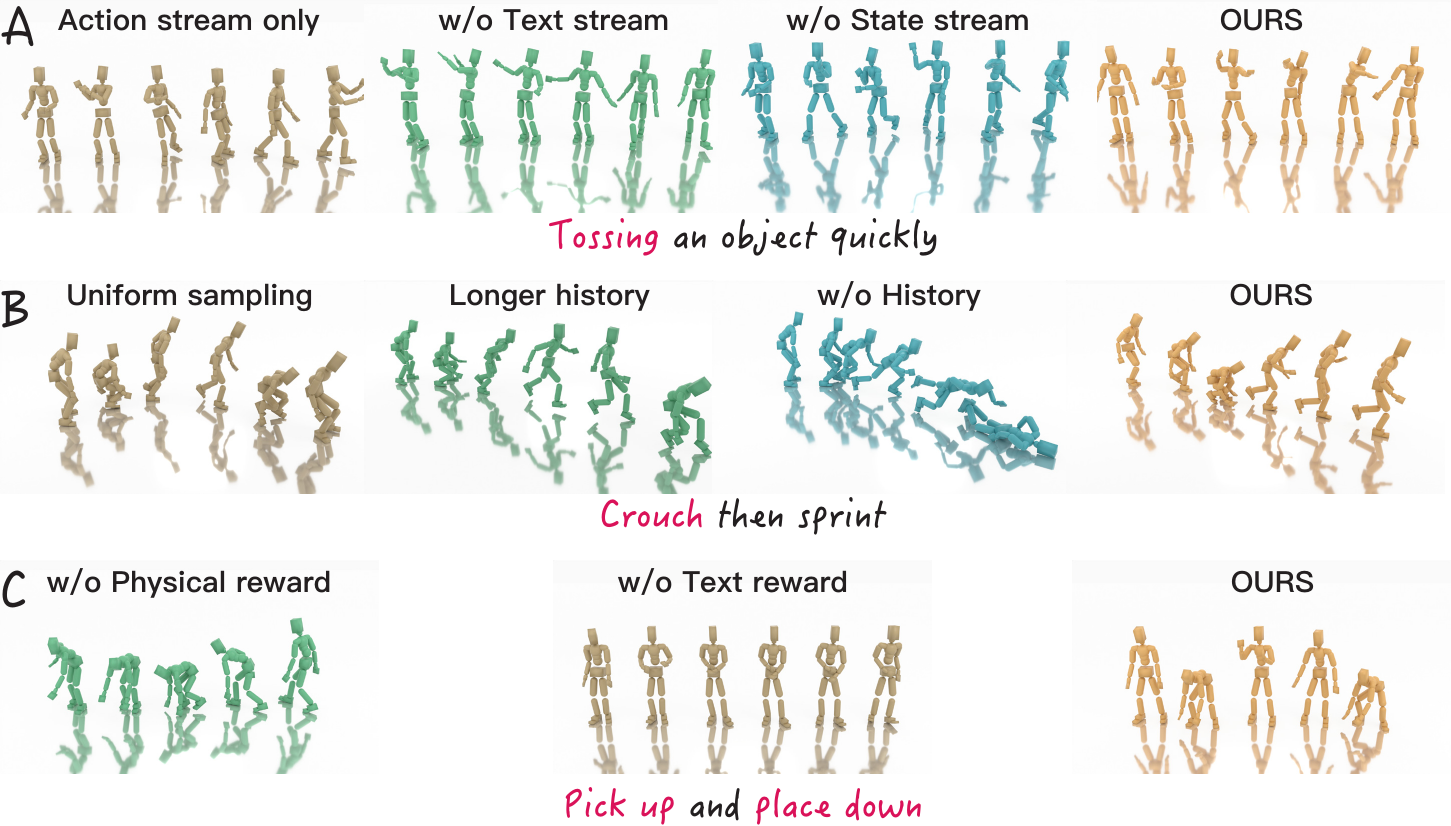}
  \caption{Qualitative ablation results. The full model preserves stable and prompt-faithful motion, while ablated variants exhibit failures in stability, prompt-following, or motion quality.}
    \Description{Rows of pose sequences for the same prompt: the full model stays stable and follows the prompt, while each ablated variant falls, drifts from the instruction, or degrades in motion quality.}
  \label{fig:ablation}
\end{figure*}
\begin{table}[t]
  \caption{Ablation studies on the HumanML3D test set.}
  \Description{Table of ablation results on the HumanML3D test set, one row per removed component, showing how each removal degrades the metrics relative to the full model.}
  \label{tab:ablation}
  \centering
  \small
  \renewcommand{\arraystretch}{1.08}
  \resizebox{\columnwidth}{!}{
  \begin{tabular}{lccc@{\hspace{8pt}}cccc}
    \toprule
    Method
    & \multicolumn{3}{c}{R-precision}
    & \multicolumn{3}{c}{Motion Quality}
    & Duration \\
    \cmidrule(lr){2-4}
    \cmidrule(lr){5-7}
    \cmidrule(lr){8-8}
    & Top-1 $\uparrow$
    & Top-2 $\uparrow$
    & Top-3 $\uparrow$
    & FID $\downarrow$
    & MM Dist. $\downarrow$
    & Div. $\rightarrow$
    & Time $\uparrow$ \\
    \midrule
    Phys-GT
    & $0.651{\pm}.002$
    & $0.815{\pm}.001$
    & $0.882{\pm}.001$
    & $0.000{\pm}.000$
    & $1.700{\pm}.001$
    & $1.494{\pm}.008$
    & $100.00\%$ \\
    Ours-Stage I
    & $\mathbf{0.429{\pm}}.003$
    & $\mathbf{0.595{\pm}}.003$
    & $\mathbf{0.689{\pm}}.001$
    & $0.203{\pm}.001$
    & $\mathbf{2.112{\pm}}.001$
    & $1.462{\pm}.007$
    & $97.67\%$ \\
    \midrule
    \multicolumn{8}{l}{\textit{Stream Ablation}} \\
    Action stream only
    & $0.357{\pm}.002$
    & $0.524{\pm}.002$
    & $0.630{\pm}.002$
    & $0.485{\pm}.000$
    & $2.268{\pm}.000$
    & $1.433{\pm}.009$
    & $97.52\%$ \\
    w/o Text stream
    & $0.307{\pm}.001$
    & $0.461{\pm}.001$
    & $0.563{\pm}.002$
    & $0.967{\pm}.000$
    & $2.390{\pm}.000$
    & $1.357{\pm}.009$
    & $\mathbf{98.47\%}$ \\
    w/o State stream
    & $0.384{\pm}.002$
    & $0.540{\pm}.002$
    & $0.633{\pm}.001$
    & $0.307{\pm}.000$
    & $2.243{\pm}.000$
    & $1.461{\pm}.006$
    & $94.57\%$ \\
    \midrule
    \multicolumn{8}{l}{\textit{History Ablation}} \\
    w/ Uniform sampling
    & $0.419{\pm}.002$
    & $0.583{\pm}.001$
    & $0.679{\pm}.002$
    & $0.302{\pm}.000$
    & $2.169{\pm}.000$
    & $1.446{\pm}.007$
    & $96.29\%$ \\
    w/ Longer history
    & $0.395{\pm}.002$
    & $0.556{\pm}.002$
    & $0.654{\pm}.001$
    & $\mathbf{0.166}{\pm}.000$
    & $2.192{\pm}.000$
    & $\mathbf{1.468}{\pm}.010$
    & $98.14\%$ \\
    w/o History
    & $0.117{\pm}.001$
    & $0.203{\pm}.001$
    & $0.278{\pm}.001$
    & $4.063{\pm}.001$
    & $2.910{\pm}.000$
    & $0.946{\pm}.009$
    & $76.68\%$ \\
    \bottomrule
  \end{tabular}
  }
\end{table}

\subsection{Benchmark Evaluation on HumanML3D}
We compare SCRIPT against three representative baselines for physics-based humanoid control. 
Table~\ref{tab:comparison} reports the quantitative results. 
SCRIPT outperforms all baselines in semantic alignment and overall motion quality, while also achieving stronger physical plausibility.
PDP and UniPhys condition their policies on a single CLIP pooled embedding, which compresses sentence-level semantics into one global vector.
This design can obscure token-level cues such as action verbs, body-part references, and modifiers, making it difficult for the policy to associate specific linguistic details with the corresponding control decisions.
As shown in Fig.~\ref{fig:humanml3d-compare} E,G, the generated behaviors often collapse to generic motion patterns, capturing only the coarse action category while ignoring directional or manner-specific constraints, such as the specified clockwise walking direction.
In contrast, SCRIPT augments the global CLIP text condition with token-level features from the penultimate layer of the CLIP text encoder, and introduces them as an explicit text stream in JAST-DiT.
Through joint attention, these fine-grained text tokens interact directly with the action and state streams, allowing the policy to associate specific linguistic cues with the corresponding physical context.
Although CLoSD improves text alignment over action-space baselines, its planner-tracker design remains vulnerable to tracking mismatch.
When the generated reference motion exceeds the tracker's capability, such as physically implausible or abrupt motions, accumulated tracking errors can lead to falls, as shown in Fig.~\ref{fig:humanml3d-compare}A,G.
SCRIPT instead predicts actions directly in simulation, reducing the mismatch between high-level motion intent and physical execution.
Stage I trains SCRIPT by imitating expert data, but closed-loop execution can still accumulate errors and shift the policy away from the expert distribution. 
Stage II mitigates this issue through on-policy rollouts in simulation, optimizing text-alignment and physical-plausibility rewards on the policy's own trajectories. 
This explains the consistent gains in FID, Diversity, and physics metrics, while stable R-precision shows no semantic degradation.
\begin{table}[t]                               
    \caption{Reward ablation studies on the HumanML3D test set. }  
    \Description{Table of reward ablation results on the HumanML3D test set, comparing metrics when individual reward terms are removed.}
    \label{tab:reward_ablation}                                    
    \centering                                                     
    \small
    \renewcommand{\arraystretch}{1.08}
    \resizebox{\columnwidth}{!}{                                   
    \begin{tabular}{l@{\hspace{6pt}}c@{\hspace{8pt}}ccc@{\hspace{8pt}}ccc}
      \toprule                                                     
      Method                                                
      & \multicolumn{1}{c}{R-prec.}                                
      & \multicolumn{3}{c}{Motion Quality}                         
      & \multicolumn{3}{c}{Physics-based Metrics} \\
      \cmidrule(lr){2-2}                                           
      \cmidrule(lr){3-5}                                           
      \cmidrule(lr){6-8}
      & Top-3 $\uparrow$                                           
      & FID $\downarrow$                                           
      & MM Dist. $\downarrow$
      & Div. $\rightarrow$                                         
      & Floating $\downarrow$                               
      & Jerk $\downarrow$
      & Duration $\uparrow$ \\                                     
      \midrule
      Phys-GT                                                      
      & $0.882{\pm}.001$                                    
      & $0.000{\pm}.000$
      & $1.700{\pm}.001$                                           
      & $1.494{\pm}.008$
      & $17.49$                                              
      & $2.941$                                                    
      & $100\%$ \\
      Ours-Stage II                                                
      & $\mathbf{0.693}{\pm}.001$                                  
      & $\mathbf{0.164}{\pm}.001$
      & $\mathbf{2.123{\pm}}.001$                                  
      & $\mathbf{1.486}{\pm}.009$
      & $17.61$                                          
      & $1.706$                                          
      & $98.08\%$ \\
      \midrule                                                     
      \multicolumn{8}{l}{\textit{Hybrid Rewards Ablation}} \\
      w/o $r_{\mathrm{phys}}$ 
      & $0.680{\pm}.001$                                       
      & $0.220{\pm}.001$
      & $2.155{\pm}.002$
      & $1.471{\pm}.008$
      & $20.793$                                       
      & $2.254$                                              
      & $93.62\%$ \\
      w/o $r_{\mathrm{text}}$ 
      & $0.649{\pm}.003$                                       
      & $0.430{\pm}.002$
      & $2.219{\pm}.003$                                                                
      & $1.425{\pm}.011$
      & $\mathbf{15.399}$                                                                
      & $\mathbf{1.169}$                                       
      & $\mathbf{98.74}\%$ \\                                     
      \bottomrule                                           
    \end{tabular}
    }
  \end{table}                                     
\subsection{Scaling on MotionMillion}
To study how SCRIPT benefits from increased model capacity, we train three variants, ranging from 0.2B to 1.2B parameters, on the MotionMillion dataset. 
Table~\ref{tab:scaling} reports the quantitative model-size scaling results. Within the evaluated range, all metrics improve consistently as model capacity increases, suggesting that SCRIPT can effectively exploit additional capacity when trained on large-scale text-motion data. Figure~\ref{fig:scaling} further shows qualitative results from our Huge model trained on MotionMillion, illustrating the diverse and physically plausible motions enabled by large-scale training.
With increased model capacity, SCRIPT learns a broader motion repertoire, covering a wide range of instructions while maintaining physically plausible control.

\subsection{Ablation Studies}

\paragraph{Effect of JAST-DiT}
Table~\ref{tab:ablation} reports the stream ablation results, validating the necessity of each modality in JAST-DiT. The Action-only variant performs poorly across all metrics, indicating that a single stream cannot capture the coupled dependencies among control, physical state, and language.
Removing the Text stream causes severe degradation in semantic alignment, as the policy loses token-level linguistic cues and tends to produce state-driven motions regardless of the prompt.
Removing the State stream preserves semantic alignment relatively well but degrades motion quality and physical stability, confirming that explicit state tokens are essential for closed-loop control.
As visualized in Figure~\ref{fig:ablation}A, these failures lead to either poor instruction following or unstable physical rollouts.
The ablation results demonstrate that the three streams play complementary roles: text carries semantic intent, state anchors physical context, and action represents executable control, while joint attention enables JAST-DiT to align language with closed-loop control.
\paragraph{Effect of Nonlinear History Conditioning.}
Table~\ref{tab:ablation} reports the history ablation results, validating the role of temporal context and the design of our nonlinear sampler.
Removing history entirely causes severe degradation, confirming that stable long-horizon closed-loop control requires sufficient temporal context.
At the same window length, replacing nonlinear sampling with uniform sampling weakens both semantic alignment and motion quality, as uniform sampling under-represents recent dynamics that are critical for immediate control.
Extending the history window further marginally improves motion quality but reduces semantic alignment, suggesting that excessive distant context can make the policy over-rely on past dynamics and respond less precisely to the input prompt.
As shown in Fig.~\ref{fig:ablation}B, these trends appear as unstable motion or weaker instruction following.

\paragraph{Effect of Hybrid Reward.}
Table~\ref{tab:reward_ablation} reports the reward ablation results, evaluating the contribution of each component in RLHR.
Without the physical reward, the policy preserves semantic alignment but loses physical plausibility, producing text-relevant motions that suffer from drift, jerky movement, or early termination in simulation.
Without the text reward, the policy improves physical metrics and may even outperform the full model on some stability-related measures, but semantic alignment and motion quality degrade.
This apparent improvement reflects a survival bias where the policy favors conservative motions to avoid early termination and maximize $r_{\mathrm{phys}}$. The trajectory-level text reward mitigates this, preserving semantic alignment.
As shown in Fig.~\ref{fig:ablation}C, these ablations produce complementary failure modes: unstable rollouts without the physical reward, and conservative motions with weak instruction following without the text reward.
The full hybrid reward avoids both extremes, enabling motions that are both faithful to language and physically plausible.
\begin{table}[t]
  \caption{Robustness under external pushes on HumanML3D.}
  \Description{Table comparing Stage I and Stage II under pushes of increasing force, reporting duration, fall rate within two seconds, FID, and R-precision at each force level.}
  \label{tab:perturbation}
  \centering
  \small
  \renewcommand{\arraystretch}{1.08}
  \resizebox{\columnwidth}{!}{
  \begin{tabular}{lcc@{\hspace{8pt}}cc@{\hspace{8pt}}cc@{\hspace{8pt}}cc}
    \toprule
    Push
    & \multicolumn{2}{c}{Fall@2s (\%) $\downarrow$}
    & \multicolumn{2}{c}{Duration time $\uparrow$}
    & \multicolumn{2}{c}{FID $\downarrow$}
    & \multicolumn{2}{c}{Top-1 $\uparrow$} \\
    \cmidrule(lr){2-3}
    \cmidrule(lr){4-5}
    \cmidrule(lr){6-7}
    \cmidrule(lr){8-9}
    & Stage I
    & Stage II
    & Stage I
    & Stage II
    & Stage I
    & Stage II
    & Stage I
    & Stage II \\
    \midrule
    0\,N
    & $1.48$
    & $\mathbf{1.38}$
    & $97.67\%$
    & $\mathbf{98.08\%}$
    & $0.203$
    & $\mathbf{0.164}$
    & $0.429$
    & $\mathbf{0.435}$ \\
    400\,N
    & $4.04$
    & $\mathbf{3.40}$
    & $96.78\%$
    & $\mathbf{97.24\%}$
    & $0.504$
    & $\mathbf{0.420}$
    & $0.359$
    & $\mathbf{0.366}$ \\
    600\,N
    & $10.42$
    & $\mathbf{9.48}$
    & $93.49\%$
    & $\mathbf{94.40\%}$
    & $0.858$
    & $\mathbf{0.727}$
    & $0.295$
    & $\mathbf{0.305}$ \\
    800\,N
    & $27.36$
    & $\mathbf{25.15}$
    & $84.84\%$
    & $\mathbf{86.95\%}$
    & $1.338$
    & $\mathbf{1.206}$
    & $0.237$
    & $\mathbf{0.244}$ \\
    \bottomrule
  \end{tabular}
  }
\end{table}
\begin{figure}[t]
    \centering
    \includegraphics[width=0.85\linewidth]{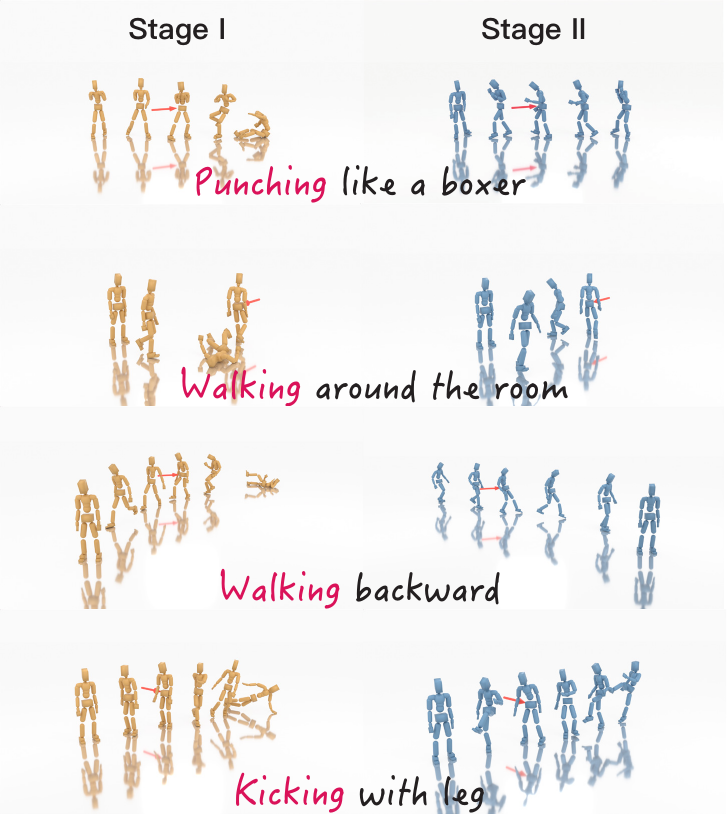}
    \caption{Qualitative comparison under an identical 600\,N pelvis push. Stage~I loses balance and falls, whereas Stage~II absorbs the disturbance and resumes the instructed motion.}
    \Description{Two rows of pose sequences under the same 600 N pelvis push: the Stage I policy loses balance and falls, while the Stage II policy recovers and continues the instructed motion.}
    \label{fig:force}
\end{figure}
\paragraph{Robustness under External Perturbations.}
To quantify the robustness gains from the post-training stage, we evaluate the models under external perturbations on the HumanML3D test split. Specifically, during each rollout, a constant horizontal force pulse is applied to the pelvis of the humanoid for 0.2\,s. The force magnitude ranges from zero for the unperturbed baseline up to 800\,N. For a strict paired comparison, we evaluate Stage~I and Stage~II using identical randomly sampled perturbation directions and onset times across all test sequences and force levels. 
As shown in Table~\ref{tab:perturbation}, the performance advantage of Stage~II becomes more pronounced as the perturbation magnitude increases.
Notably, at every nonzero force level, Stage~II consistently achieves higher Duration time and a lower Fall@2s metric, which we define as the fall rate within 2\,s after the push begins. 
Stage~II also maintains better alignment with language instructions under external force. It consistently yields lower FID and higher R-precision under stronger impacts. These results demonstrate that the RL optimization stage does not merely induce conservative behavior. Rather, Stage~II successfully improves physical robustness while preserving the intended motion semantics.
Qualitative comparisons under identical pushes are shown in Fig.~\ref{fig:force}.

\section{Conclusion}
In this paper we present SCRIPT, a scalable diffusion policy framework for language-driven physics-based humanoid control. By jointly modeling actions, physical states, and language with JAST-DiT, and further incorporating history conditioning and reinforcement learning post-training, SCRIPT learns closed-loop control policies from large-scale physically executable trajectories. Experiments show that SCRIPT outperforms prior motion-generation and planner--tracker methods in instruction following, motion quality, and physical plausibility, while exhibiting consistent gains with data and model scaling. Future work will extend SCRIPT to more complex human-object interactions, multi-agent collaboration, and open-environment tasks. 
We discuss limitations and failure modes in the supplementary material.

\begin{acks}
This work was supported in part by the National Natural Science Foundation of China under Grant No.~W2431046; the Central Guidance on Local Science and Technology Development Fund of China under Grant No.~YDZX20253100001001; the Key Laboratory of Intelligent Perception and Human-Machine Collaboration (ShanghaiTech University), Ministry of Education; and the Shanghai Frontiers Science Center of Human-Centered Artificial Intelligence. Also supported by HPC Platform of ShanghaiTech University.
\end{acks}
\bibliographystyle{ACM-Reference-Format}
\bibliography{reference}
\clearpage
\appendix
\graphicspath{{./}{supp/}}   
\section{Data Curation and Processing Details}
\label{supp:Data}
This section provides implementation details of our data curation procedure, including artifact filtering,  perturbed rollouts for robustness checking, and padding and window slicing of the tracked trajectories.
\subsection{Preprocessing and Filtering}
\label{supp:data_filter}
We organize all motion clips following the MotionMillion data format~\cite{fan2025go}. 
For each SMPL sequence, we compute the 3D positions of 24 joints at each frame via forward kinematics, denoted as $p_t^{(j)} \in \mathbb{R}^3$. 
We use these joint positions for geometric filtering and rollout quality assessment, filtering both the original kinematic motions and simulated rollouts to obtain high-quality references and tracking data.
For kinematic motions, we filter clips using four criteria: sequence length, motion magnitude, ground penetration, and floating artifacts.
\begin{enumerate}
    \item \textbf{Short clips.} We remove clips with $T < 30$ frames, since their temporal span is insufficient to form valid history and prediction windows.
    
    \item \textbf{Near-static clips.} We compute the mean joint-angle difference within a 1-second sliding window. If the mean difference remains below $2\times10^{-3}$ for all windows, the clip is considered to contain insufficient motion and is discarded.
    
    \item \textbf{Ground penetration.} We define $z_t=\min_j p_{t,z}^{(j)}$ as the lowest joint height at frame $t$. A clip is marked as penetrating if the median value of $z_t$ over the final 1 second is more than $0.05\,\mathrm{m}$ lower than that over the first 1 second, and the minimum value of $z_t$ over the full sequence is below $-0.03\,\mathrm{m}$.
    
    \item \textbf{Floating artifacts.} If the minimum foot-joint height remains above $0.30\,\mathrm{m}$ throughout the sequence and its variance is below $0.08$, we treat the clip as lacking valid ground contact and remove it.
\end{enumerate}
After raw-motion filtering, some kinematic references may still violate rigid-body constraints when tracked in simulation, leading to tracking failures or jerk artifacts. 
We therefore further filter simulated rollouts using root-aligned MPJPE to measure tracking accuracy and jerk to measure rollout smoothness.
Let $p_t^{(j)}$ and $\hat{p}_t^{(j)}$ denote the reference and rollout joint positions, respectively, with the pelvis joint indexed by $j=0$. We compute root-aligned MPJPE as
\begin{equation}
\overline{\mathrm{MPJPE}}
=
\frac{1}{TJ}
\sum_{t=1}^{T}
\sum_{j=1}^{J}
\left\|
(\hat p_t^{(j)}-\hat p_t^{(0)})
-
(p_t^{(j)}-p_t^{(0)})
\right\|_2 .
\label{eq:mpjpe_filter}
\end{equation}

For rollout smoothness, we compute jerk using the third-order finite difference of joint positions with $\Delta t=1/\mathrm{fps}$:
\begin{equation}
\overline{\mathrm{jerk}}
=
\frac{1}{(T-3)J}
\sum_{t=1}^{T-3}
\sum_{j=1}^{J}
\left\|
\frac{
\hat p_{t+3}^{(j)}
-3\hat p_{t+2}^{(j)}
+3\hat p_{t+1}^{(j)}
-\hat p_t^{(j)}
}{\Delta t^3}
\right\|_2 .
\label{eq:jerk_filter}
\end{equation}
\begin{figure}[t]
    \centering
    \includegraphics[width=\linewidth]{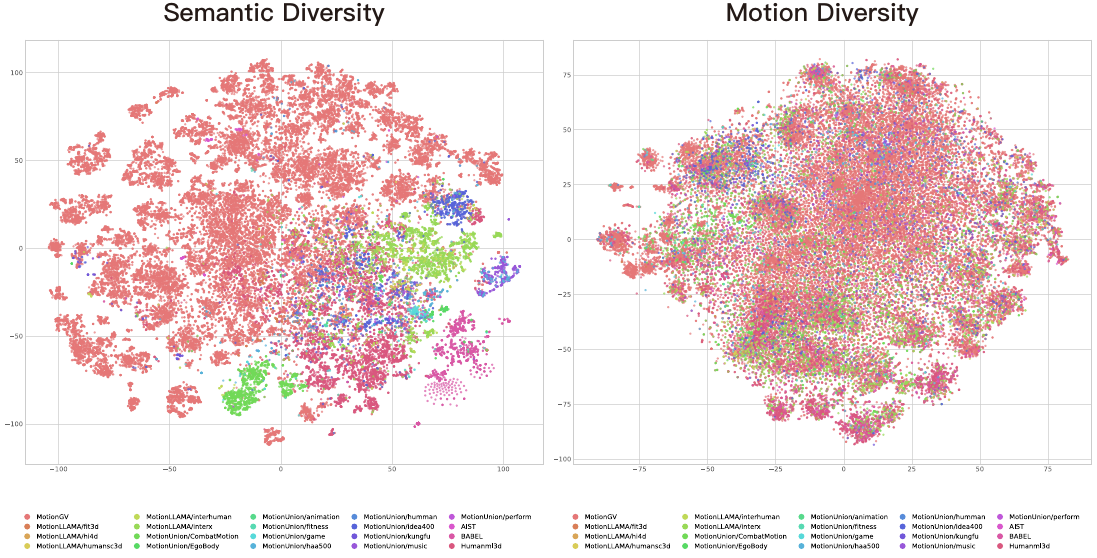}
    \caption{t-SNE visualization of the retained data. }
    \Description{Two t-SNE scatter plots: left, rollout state embeddings; right, CLIP text embeddings of the retained clips, both spread broadly without dominant clusters.}
    \label{fig:diversity}
\end{figure}
We discard rollouts with $\overline{\mathrm{MPJPE}} \geq 0.15\,\mathrm{m}$ or $\overline{\mathrm{jerk}} \geq 600\,\mathrm{m/s^3}$ to remove severe tracking failures and high-frequency artifacts.
\subsection{Perturbed Rollout}
\label{supp:perturbed_rollout}

To efficiently process large-scale motion data, we divide the pre-filtered motion clips into multiple subsets and train a separate tracking policy~\cite{luo2023perpetual} for each subset. 
After training, each policy tracks the reference motions from its corresponding subset in parallel, accelerating simulation tracking and state-action data collection. 
For each rollout, we record the proprioceptive state $\mathbf{s}_t$ from the simulator and the target PD control action $\mathbf{a}_t$ predicted by the tracking policy.
Specifically, we partition the pre-filtered clips into 160 subsets of approximately 7.2K clips each.
Each subset is tracked by a dedicated expert policy trained in 512 parallel environments, following the default PHC configuration.
Training concludes when the success rate exceeds 90\% for five consecutive evaluations, where success is defined as completing rollouts without falling or large tracking deviations.
Following PDP~\cite{truong2024pdp}, we inject small Gaussian perturbations into the executed tracking actions during rollout to broaden the collected state distribution. 
Behavior cloning on clean expert trajectories mainly covers states near the expert distribution, whereas closed-loop execution may drift away from this distribution due to accumulated prediction errors. 
Action perturbations expose the tracking policy to off-reference states, while the unperturbed policy output provides corrective supervision for recovery. 
This produces perturbed-state/clean-action pairs that improve robustness to state deviations.

Specifically, at simulation step $t$, the tracking policy outputs an unperturbed action $\mathbf{a}_t$. 
We add isotropic Gaussian noise to obtain the executed action:
\begin{equation}
\tilde{\mathbf{a}}_t
=
\mathbf{a}_t + \boldsymbol{\epsilon}_t,
\qquad
\boldsymbol{\epsilon}_t \sim \mathcal{N}(\mathbf{0}, \sigma^2\mathbf{I}),
\qquad
\sigma = 0.01.
\label{eq:perturbed_action}
\end{equation}
The perturbed action $\tilde{\mathbf{a}}_t$ is sent to the PD controller, while the unperturbed action $\mathbf{a}_t$ is saved as supervision. 
Thus, each training pair consists of the perturbed rollout state $\mathbf{s}_t^{\mathrm{pert}}$ and the corresponding unperturbed action $\mathbf{a}_t$. 
This encourages the downstream diffusion policy to recover from off-reference states during closed-loop execution.

\subsection{Data Distribution Visualization}
To examine the coverage of the retained data, we visualize both the language and motion distributions using t-SNE~\cite{van2008visualizing}, as shown in Fig.~\ref{fig:diversity}. 
We represent motion with rollout proprioceptive states $\mathbf{s}_t$ and language with CLIP pooled text embeddings $\mathbf{c}_{\mathrm{pool}}$, and apply t-SNE to visualize motion diversity and semantic coverage.
The state and text visualizations provide complementary views of the collected data. 
State-space embeddings reveal the diversity of tracked physical motion patterns, while text embeddings reflect the semantic coverage of the language conditions.
Together, they suggest that the retained physically executable data preserve both motion diversity and semantic richness.
\subsection{Windowed Sample Construction}
\label{supp:window_construction}
To construct fixed-length training samples, we apply first-frame padding and sliding-window slicing to each tracked trajectory. 
Since nonlinear history sampling requires sufficient past context, we prepend $N_{\mathrm{pad}}$ copies of the first frame to support boundary samples near the beginning of a sequence. 
We then slice the padded trajectory into fixed-length windows according to the history range and prediction horizon $H$. 
Each window contains the historical states for constructing $\mathcal{H}_t$, the subsequent $H$-frame state-action supervision targets, and the corresponding text condition $c$.
\section{Online Reinforcement Learning Post-Training}
\label{supp:rl_post_training}
\subsection{Stochastic Flow Sampling}
\label{supp:stochastic_flow_sampling}
Following ReinFlow~\cite{zhang2025reinflow}, we compute policy likelihoods by treating the stochastic flow sampling process as a Markov chain over Euler steps.
Let $\mathbf{x}_k$ denote the intermediate noisy state-action trajectory chunk at the $k$-th Euler step.
The deterministic part of the Euler update defines the transition mean
\begin{equation}
\boldsymbol{\mu}_k
=
\mathbf{x}_k
+
v_\theta(\mathbf{x}_k,\tau_k,\mathbf{y})\Delta\tau,
\label{eq:supp_euler_mean}
\end{equation}
where $\mathbf{y}=\{\mathcal{H}_t,\mathbf{c}_{\mathrm{txt}},\mathbf{c}_{\mathrm{pool}}\}$ denotes the combined policy condition.
Let $P_a(\cdot)$ denote the projection operator that extracts the action dimensions from the full state-action chunk.  
We define
$\mathbf{x}_k^a=P_a(\mathbf{x}_k)$,
$\boldsymbol{\mu}_k^a=P_a(\boldsymbol{\mu}_k)$, and
$\boldsymbol{\sigma}_k=\sigma_\phi(\tau_k,\mathcal{H}_t)$,
where $\boldsymbol{\sigma}_k\in\mathbb{R}^{|\mathbf{a}|}$ is the per-dimension exploration scale restricted to the action subspace.  
Since stochasticity is injected exclusively into the action dimensions, the chain log-likelihood is evaluated as: 
\begin{equation}
\begin{aligned}
\log\pi_\theta(\mathbf{x}_{0:K}\mid\mathbf{y})
&=
\log\mathcal{N}(\mathbf{x}_0;\mathbf{0},\mathbf{I})+
\sum_{k=0}^{K-1}
\log\mathcal{N}
\left(
\mathbf{x}^a_{k+1};
\boldsymbol{\mu}^a_k,
\mathrm{diag}(\boldsymbol{\sigma}_k^2)
\right).
\end{aligned}
\label{eq:supp_chain_logprob}
\end{equation}
The initial noise term $\log\mathcal{N}(\mathbf{x}_0;\mathbf{0},\mathbf{I})$ is independent of the policy parameters and safely cancels out in the PPO likelihood ratio calculations.  
Computing likelihoods only in the action subspace avoids treating deterministic future-state predictions as policy entropy, while retaining tractable transition likelihoods for policy-gradient optimization.  

\subsection{Exploration Noise Network}
\label{supp:noise_network}
We further detail the lightweight network used to predict the exploration scale 
$\sigma_\phi(\tau_k,\mathcal{H}_t)$ in Eq.9. 
Consistent with the main text, the network takes the flow time $\tau_k$ and the sampled state history $\mathcal{H}_t$ as inputs, rather than the AdaLN-Zero condition used by JAST-DiT. 
Specifically, we first compute a raw output
\[
\mathbf{z}_k
=
g_\phi\!\left([\mathrm{emb}(\tau_k); h(\mathcal{H}_t)]\right),
\]
where $\mathrm{emb}(\tau_k)$ is a sinusoidal time embedding, $h(\mathcal{H}_t)$ is a history projection, and $g_\phi$ is a lightweight MLP.
To keep the exploration scale bounded, we map $\mathbf{z}_k$ to the log-variance space:
\begin{equation}
\log \boldsymbol{\sigma}_k^2
=
\log\sigma_{\min}^2
+
\left(\log\sigma_{\max}^2-\log\sigma_{\min}^2\right)
\frac{1+\tanh(\mathbf{z}_k)}{2}.
\label{eq:supp_logvar_bound}
\end{equation}
The final exploration scale is 
$\boldsymbol{\sigma}_k=\sigma_\phi(\tau_k,\mathcal{H}_t)
=\exp(\frac{1}{2}\log \boldsymbol{\sigma}_k^2)$.
This parameterization constrains each action-dimension standard deviation to $[\sigma_{\min},\sigma_{\max}]$, preventing the exploration scale from collapsing to zero or growing unbounded during PPO updates.
\subsection{Optimization Details}
\label{supp:optimization_objective}
In this section, we elaborate on the three core components of the RLHR objective: the chain-level PPO ratio, the closed-form entropy bonus, and the velocity-field BC anchor.
Based on the chain log-likelihood derived in Eq.~\ref{eq:supp_chain_logprob}, the PPO importance ratio is evaluated over the entire stochastic Euler denoising chain.

\paragraph{Chain-level PPO Ratio.}
Let the per-step action-space transition log-density be
\begin{equation}
\log\pi_{\theta,\phi}^{(k)}
=
\log\mathcal{N}
\left(
\mathbf{x}_{k+1}^{a};
\boldsymbol{\mu}_{k}^{a},
\mathrm{diag}(\boldsymbol{\sigma}_k^2)
\right),
\label{eq:supp_step_logprob}
\end{equation}
where $\mathbf{x}_{k+1}^{a}$ and $\boldsymbol{\mu}_{k}^{a}$ denote the action-dimension sample and transition mean, respectively, and $\boldsymbol{\sigma}_k$ is the corresponding exploration scale.  
Because the initial noise term in Eq.~\ref{eq:supp_chain_logprob} is independent of the policy parameters, it safely cancels out when computing the importance sampling ratio $\rho(\theta,\phi)$ between the current and old policies: 
\begin{equation}
\rho(\theta,\phi)
=
\exp\!\left(
\sum_{k=0}^{K-1}
\left[
\log\pi_{\theta,\phi}^{(k)}
-
\log\pi_{\theta_{\mathrm{old}},\phi_{\mathrm{old}}}^{(k)}
\right]
\right).
\label{eq:supp_chain_ratio}
\end{equation}
The clipped PPO surrogate objective is thus formulated as:
\begin{equation}
\mathcal{L}_{\mathrm{PPO}}
=
-\mathbb{E}
\left[
\min
\left(
\rho \hat{A},
\mathrm{clip}(\rho,1-\epsilon,1+\epsilon)\hat{A}
\right)
\right],
\label{eq:supp_ppo_clip}
\end{equation}
where the advantage estimate $\hat{A}$ is computed using GAE from the physics rollouts.  
In practice, we clamp the accumulated log-probability differences before exponentiation to prevent numerical overflow, which can otherwise easily occur when summing over high-dimensional action spaces across multiple Euler integration steps.

\paragraph{Closed-form Gaussian Entropy.}
Because each stochastic Euler transition is modeled as an independent diagonal Gaussian within the action subspace, its step-wise entropy admits a closed-form expression:
\begin{equation}
    H_k(\phi) 
    = 
    \frac{1}{2} 
    \sum_{i=1}^{|\mathbf{a}|} 
    \left[ 
    \log(2\pi e) 
    + 
    2\log\boldsymbol{\sigma}_{k,i} 
    \right],
    \label{eq:supp_per_step_entropy}
\end{equation}
where $\boldsymbol{\sigma}_{k,i}$ denotes the predicted standard deviation for the $i$-th action dimension at step $k$. 
The overall policy entropy bonus is then computed as the average over the entire integration chain:
\begin{equation}
    \mathcal{H}_{\mathrm{entropy}} 
    = 
    \frac{1}{K} 
    \sum_{k=0}^{K-1} 
    H_k(\phi).
    \label{eq:supp_entropy}
\end{equation}
Crucially, this formulation is exactly differentiable with respect to the noise network parameters $\phi$ via $\boldsymbol{\sigma}_k=\sigma_\phi(\tau_k,\mathcal{H}_t)$. This property allows us to systematically regularize the exploration scale without relying on high-variance, sampling-based entropy estimators.

\paragraph{Velocity-field BC Anchor.}
The BC anchor introduced in Eq.13 is designed as a velocity-field regularizer rather than a direct action-space constraint.  
By anchoring the fine-tuned velocity field $v_\theta$ to the frozen pretrained model $v_{\theta_0}$, this term preserves the foundational geometric prior established during Stage-I flow matching, while allowing PPO to optimize the induced action distribution for specific task rewards.  
To maintain strict consistency with the original flow-matching objective, the BC triplets $(\mathbf{x},\tau,\boldsymbol{\epsilon})$ are independently resampled for each update rather than reusing the integration chains from physics rollouts. This decoupled sampling ensures that the anchor remains a robust regularizer, preventing the policy from collapsing toward sub-optimal local modes during reinforcement learning.
\section{Text-State Aligner}
\label{supp:Aligner}

To establish a shared representation space between physical state trajectories and language instructions, we train a text-state contrastive aligner.
This aligner remains frozen during the RLHR stage and is utilized for both semantic reward computation and evaluation. 
The model follows a symmetric two-tower architecture comprising a state trajectory encoder and a text encoder, which map their respective modalities into a shared 512-dimensional latent space. 
Each training sample contains a state sequence $\mathbf{S}=\{\mathbf{s}_t\}_{t=1}^{T}$, where $\mathbf{s}_t\in\mathbb{R}^{358}$, and the CLIP pooled embedding of the corresponding text, $\mathbf{c}_{\mathrm{pool}}\in\mathbb{R}^{1280}$~\cite{radford2021learning}.
The CLIP embeddings are extracted offline with a frozen text encoder and cached before training.
The state inputs are normalized to $[-1,1]$ with a min--max normalizer before encoding.

The state encoder first projects each 358-dimensional state vector to a 1024-dimensional latent space with a linear layer.
We prepend a learnable query token to the sequence, add sinusoidal positional encodings, and feed the resulting tokens into an 8-layer Transformer encoder~\cite{vaswani2017attention} with hidden dimension 1024, FFN dimension 2048, 8 attention heads, and dropout 0.1.
The output corresponding to the query token is then projected to 512 dimensions and L2-normalized to obtain the state embedding $\mathbf{z}_s$.
The text encoder consists of a linear projection that maps the precomputed CLIP pooled embedding $\mathbf{c}_{\mathrm{pool}}$ to the same 512-dimensional space, followed by L2 normalization to obtain the text embedding $\mathbf{z}_t$.

We train the two-tower model with a symmetric InfoNCE loss~\cite{oord2018representation}.
For a batch of $B$ paired state sequences and text conditions, let $\mathbf{z}_s^{(i)}$ and $\mathbf{z}_t^{(j)}$ denote the normalized embeddings of the $i$-th state sequence and the $j$-th text condition, respectively.
The similarity matrix is defined as
\begin{equation}
S_{ij}
=
\gamma \cdot
\mathbf{z}_s^{(i)\top}
\mathbf{z}_t^{(j)},
\label{eq:supp_aligner_similarity}
\end{equation}
where $\gamma$ is a learnable temperature parameter.
The training objective averages the state-to-text and text-to-state cross-entropy losses:
\begin{equation}
\mathcal{L}_{\mathrm{align}}
=
\frac{1}{2}
\left[
\mathrm{CE}(S,\mathbf{y})
+
\mathrm{CE}(S^\top,\mathbf{y})
\right],
\qquad
\mathbf{y}=(1,\ldots,B).
\label{eq:supp_aligner_loss}
\end{equation}

We train the aligner with AdamW for 300 epochs, using an initial learning rate of $1\times10^{-4}$, weight decay $1\times10^{-4}$, and batch size 96.
The learning rate is decayed to $1\times10^{-7}$ with cosine annealing.
The maximum state sequence length is set to 400, matching the pre-sliced training window size.
After training, the aligner is frozen and used to extract state and text features, as well as to score the semantic consistency between generated state sequences and language conditions.
\section{Evaluation Metrics}
\label{supp:Metric}

\paragraph{Text-following Metrics.}
We compute text-following metrics in the shared embedding space of the frozen text-state aligner described in Sec.~\ref{supp:Aligner}. 
Given $B$ paired generated state sequences and text conditions, let $\mathbf{z}_s^{(i)}$ and $\mathbf{z}_t^{(i)}$ denote their normalized state and text embeddings.
For R-Precision, we retrieve the paired text for each generated state sequence within the batch:
\begin{equation}
\mathrm{R@}k
=
\frac{1}{B}
\sum_{i=1}^{B}
\mathbb{I}
\left[
i \in
\mathrm{TopK}_k
\left(
\left\{
\mathbf{z}_s^{(i)\top}\mathbf{z}_t^{(j)}
\right\}_{j=1}^{B}
\right)
\right].
\label{eq:supp_r_precision}
\end{equation}
We report $k=1,2,3$, where higher values indicate better text--motion alignment.

MM-Dist measures the average distance between paired state and text embeddings:
\begin{equation}
\mathrm{MM\text{-}Dist}
=
\frac{1}{B}
\sum_{i=1}^{B}
\left\|
\mathbf{z}_s^{(i)}-\mathbf{z}_t^{(i)}
\right\|_2 .
\label{eq:supp_mm_dist}
\end{equation}
Lower values indicate closer semantic alignment.

For FID, we fit Gaussian distributions to the generated and reference state embeddings, denoted by $(\boldsymbol{\mu}_{\mathrm{gen}},\boldsymbol{\Sigma}_{\mathrm{gen}})$ and $(\boldsymbol{\mu}_{\mathrm{ref}},\boldsymbol{\Sigma}_{\mathrm{ref}})$, respectively:
\begin{equation}
\mathrm{FID}
=
\left\|
\boldsymbol{\mu}_{\mathrm{gen}}
-
\boldsymbol{\mu}_{\mathrm{ref}}
\right\|_2^2
+
\mathrm{Tr}
\left(
\boldsymbol{\Sigma}_{\mathrm{gen}}
+
\boldsymbol{\Sigma}_{\mathrm{ref}}
-
2
\left(
\boldsymbol{\Sigma}_{\mathrm{gen}}
\boldsymbol{\Sigma}_{\mathrm{ref}}
\right)^{1/2}
\right).
\label{eq:supp_fid}
\end{equation}
Lower FID indicates that the generated state distribution is closer to the reference distribution.

\paragraph{Physical-plausibility Metrics.}
We compute physical-plausibility metrics directly from the global SMPL joint positions $\mathbf{p}_t^{(j)}\in\mathbb{R}^3$, where $j=1,\ldots,J$ and the $z$ axis points upward.
Floating measures the average positive clearance between the lowest body joint and the ground, with a tolerance $\tau_{\mathrm{tol}}=0.005\,\mathrm{m}$:
\begin{equation}
\mathrm{Floating}
=
\frac{10^3}{T}
\sum_{t=1}^{T}
\max
\left(
\min_j p_{t,z}^{(j)}
-
\tau_{\mathrm{tol}},
0
\right)
\quad [\mathrm{mm}].
\label{eq:supp_floating}
\end{equation}

Jerk measures trajectory smoothness using the same third-order finite-difference form as Eq.~\eqref{eq:jerk_filter}, but for evaluation we report the frame-based value in $\mathrm{mm/frame^3}$. 
Specifically, we omit the $\Delta t^3$ normalization and multiply by $10^3$ to convert meters to millimeters.
Lower values indicate smoother motions with fewer high-frequency artifacts.

Duration Time measures how long the agent remains valid during simulation before falling.
Following the early-termination criterion used in PHC, we terminate a rollout when the pelvis height falls below a threshold, i.e., $p_{t,z}^{(\mathrm{root})}<\tau_{\mathrm{fall}}$.
If no fall is detected, the rollout is considered successfully tracked to the end.
We define Duration Time as the ratio between the total number of valid simulated frames before termination and the total number of reference frames across all rollouts:
\begin{equation}
\mathrm{Duration}
=
\frac{
\sum_{n=1}^{N} T_{\mathrm{valid}}^{(n)}
}{
\sum_{n=1}^{N} T_{\mathrm{ref}}^{(n)}
}.
\label{eq:supp_duration}
\end{equation}
We set $\tau_{\mathrm{fall}}=0.15\,\mathrm{m}$, consistent with the default termination height in PHC, and use the same threshold for all methods.
\section{Training Details}
\label{supp:training_details}

\paragraph{HumanML3D Pretraining.}
We train SCRIPT-Large, a 0.5B-scale model with 577.97M parameters, on HumanML3D as the main pretrained model for benchmark comparison.
The model uses a 10-layer JAST-DiT with hidden dimension 768, 6 attention heads, and head dimension 128.
We set the prediction horizon to $H=4$ and the history length to $L_{\max}=154$, resulting in a training window of 158 frames with stride 2.

Training is performed on a single node with 8 A100 GPUs using PyTorch DDP.
We use bf16 mixed precision and enable TF32 acceleration.
The optimizer is AdamW with learning rate $1\times10^{-4}$, a 1K-step linear warmup, $\beta=(0.9,0.999)$, and weight decay $1\times10^{-4}$.
We use an EMA decay of 0.9999, enabled after the first 1K warmup steps.
The per-GPU batch size is 1024, giving an effective global batch size of 8192.
The model is trained for 110K steps.
For classifier-free guidance, the text condition is replaced with an empty string with probability 10\%.
We use the EMA weights of the pretrained model as the behavior-cloning checkpoint for downstream RL initialization.

\paragraph{MotionMillion Scaling Pretraining.}
For scaling experiments on MotionMillion, we increase the history length to $L_{\max}=604$ and use a training window of 608 frames with stride 2.
We train three model variants that differ only in network depth and width: Base, Large, and Huge.
Base uses 8 layers, hidden dimension 512, and 4 attention heads; Large uses 10 layers, hidden dimension 768, and 6 attention heads; Huge uses 12 layers, hidden dimension 1024, and 8 attention heads.
The head dimension is fixed to 128 for all variants.
These models have approximately 0.2B, 0.5B, and 1.2B parameters, respectively.

All variants are trained on a single node with 8 A100 GPUs using PyTorch DDP and bf16 mixed precision.
We use AdamW with learning rate $1\times10^{-4}$ and a 1K-step warmup, together with EMA decay 0.9999.
All models are trained for 560K steps, with checkpoints saved every 20K steps.
The classifier-free guidance setting follows the HumanML3D pretraining stage, using 10\% text-condition dropout.

\paragraph{RLHR Post-Training.}
We initialize the actor from the EMA weights of the HumanML3D pretrained checkpoint and use a lightweight MLP critic with approximately 1.4M parameters.
RL post-training is performed on a single A100 GPU with 128 parallel Isaac Gym environments.
The simulation uses the PHC humanoid with 24 SMPL body parts and PD control over 23 actuated joints.

Each PPO iteration collects 300 frames per environment.
We optimize for 5 epochs with mini-batch size 1280 and train for up to 300 iterations.
Both actor and critic use cosine learning-rate schedules, with peak learning rates of $1\times10^{-5}$ and $1\times10^{-3}$, respectively, and a first cycle length of 1500 optimization steps.
We use PPO clipping threshold $\epsilon=0.10$, discount factor $\gamma=0.99$, maximum gradient norm 1.0, and running reward scaling.

During Euler sampling, we inject learnable Gaussian exploration noise only into the action dimensions.
The noise scale is predicted by a lightweight network and constrained to $\sigma\in[0.03,0.08]$.
The training objective combines the PPO loss, value loss, entropy regularization, and a velocity-form behavior-cloning anchor with coefficient 1.0.
The hybrid reward consists of a dense physical reward with weight 1.0 and a sequence-level episodic contrastive text reward with coefficient 20.
\paragraph{Training and Inference Cost.}
For the main HumanML3D benchmark, Stage-I pretraining of SCRIPT-Large on HumanML3D takes approximately 30 hours on a single node with eight A100 GPUs, with a measured throughput of 1.04 steps/s.
Stage-II RLHR post-training is performed on a single A100 GPU with 128 parallel simulation environments.
The checkpoint used for the reported Stage-II results was obtained after 160 PPO iterations, requiring approximately four days of training.
At inference time, SCRIPT-Large generates an action chunk of $H=4$ frames in approximately 333\,ms on a single A100 GPU.
Each prediction uses 10 denoising steps, where each step requires one conditional and one unconditional network evaluation for classifier-free guidance, resulting in 20 network evaluations in total.
Thus, the inference latency is approximately $2.5\times$ the duration of the generated action chunk.
\section{Limitations and Future Work}
\label{supp:limitations}
Although SCRIPT jointly integrates action, state, and language representations, several limitations remain.
SCRIPT currently focuses on whole-body humanoid control on flat terrain.
Therefore, SCRIPT does not support fine-grained hand articulation and grasping~\cite{luo2024omnigrasp}, humanoid loco-manipulation~\cite{he2025viral}, or scene-aware interaction in cluttered environments~\cite{pan2025tokenhsi}.
Furthermore, during data curation, we use multiple expert tracking policies in parallel to extract state-action demonstrations and retain only reference motions and simulated rollouts that pass our predefined motion-quality, tracking-error, and smoothness filters.
Although this filtering improves the physical validity of the retained data, it may underrepresent challenging yet physically feasible motions and bias the learned policy toward the curated training distribution.
Consequently, prompts describing underrepresented or out-of-distribution motions may lead to loss of balance or visible artifacts, whereas physically infeasible requests, such as sustained unsupported flight (e.g., ``flying like an airplane''), cannot be reliably realized by the current embodiment.
Finally, the current implementation is not yet optimized for high-frequency, real-time closed-loop humanoid control.
Benefiting from the flow-matching formulation, our policy is able to generate an action chunk with only 10 denoising steps.
However, further optimization is still required to achieve high-frequency real-time interactive control.
Future work will focus on improving inference efficiency through acceleration techniques, such as policy distillation~\cite{wang2024one}, reduced sampling steps~\cite{salimans2022progressive, song2023consistency}, and optimized noise schedules~\cite{ye2026world}, while extending SCRIPT toward richer human-object interactions and real-world humanoid deployment.

\section{Discussion}
\label{supp:discussion}
The JAST-DiT architecture achieves excellent performance during the initial imitation learning stage. 
Ablation experiments reveal that the history conditioning mechanism dictates generation quality and stability, while the independent text and state streams anchor semantics and physical feasibility. 
This confirms that coupling dedicated token streams through joint attention ensures accurate instruction execution and prevents motion homogenization.
Beyond the architecture, large-scale curated data underpins these gains, and all metrics improve monotonically with model capacity without saturation.
RLHR builds upon this imitation anchor to mitigate state drift through interactive exploration in physics simulation. 
This optimization grants the policy robust recovery capabilities against sudden external forces, enabling steady execution when facing perturbations.  During RL exploration, policies tend to converge toward conservative survival behaviors to avoid falling, generating degenerate motions that deviate from textual instructions.
The hybrid reward mechanism counteracts this tendency via mandatory semantic alignment, suggesting that closed-loop control systems must establish strict cross-modal constraints to balance physical stability and instruction fidelity.

\end{document}